\documentclass[aps,prl,letterpaper,superscriptaddress,showpacs,amsmath,floats,twocolumn]{revtex4-1}

\usepackage{graphicx}
\usepackage{amssymb,amsfonts,amsmath}
\usepackage{physics}
\usepackage{siunitx}
\usepackage{bm}
\usepackage[utf8]{inputenc}
\usepackage[T1]{fontenc}
\usepackage{lmodern}
\usepackage{mathtools}
\usepackage{alltt}
\usepackage{mathrsfs}
\usepackage[colorlinks]{hyperref}

\begin{document}

\title{Enhancing Sensitivity of an Atom Interferometer to the Heisenberg Limit using Increased Quantum Noise}

\author{Renpeng Fang}
\email{renpengfang2016@u.northwestern.edu}
\affiliation{Department of Physics and Astronomy, Northwestern University, 2145 Sheridan Road, Evanston, IL 60208, USA}

\author{Resham Sarkar}
\affiliation{Department of Physics and Astronomy, Northwestern University, 2145 Sheridan Road, Evanston, IL 60208, USA}

\author{Selim M. Shahriar}
\affiliation{Department of Physics and Astronomy, Northwestern University, 2145 Sheridan Road, Evanston, IL 60208, USA}
\affiliation{Department of ECE, Northwestern University, 2145 Sheridan Road, Evanston, IL 60208, USA}

\date{\today}

\begin{abstract}
In a conventional atomic interferometer employing $N$ atoms, the phase sensitivity is at the standard quantum limit: $1/\sqrt{N}$. Using spin-squeezing, the sensitivity can be increased, either by lowering the quantum noise or via phase amplification, or a combination thereof. Here, we show how to increase the sensitivity, to the Heisenberg limit of $1/N$, while increasing the quantum noise by $\sqrt{N}$, thereby suppressing by the same factor the effect of excess noise. The protocol uses a Schr\"odinger Cat state representing a superposition of two collective states of $N$ atoms, behaving as a single entity with an $N$-fold increase in Compton frequency. The resulting $N$-fold phase magnification is revealed by using atomic state detection instead of collective state detection. We also show how to realize an atomic clock based on such a Schr\"odinger Cat state, with an $N$-fold increase in the effective transition frequency. We also show how the signals, for a given protocol, produces drastically different signals for different parities of $N$, for both detection methods.  For a system that produces both odd and even values of $N$ with equal probability, we show that averaging over many instances filters out the signal for one parity, thus allowing a sensitivity that is within a factor of $\sqrt{2}$ of the Heisenberg limit, while maintaining the robustness against excess noise.  We discuss potential experimental constraints for implementing this scheme, using one axis twist squeezing employing the cavity feedback scheme, and show that the effects of cavity decay and spontaneous emission are highly suppressed. We find that the maximum improvement in sensitivity can be close to the ideal limit, for as many as ten million atoms.

OCIS Codes: 020.1335, 270.2500, 120.3940, 120.3180
\end{abstract}


\maketitle


\section{Introduction}
\label{Introduction}

In an atomic interferometer, the signal $S$ can be expressed as a function of the phase difference $\phi$ between the two arms. The measurement sensitivity, $\Lambda$, can be expressed as the inverse of the phase fluctuation (PF): $\Lambda=\text{PF}^{-1}=\abs{\partial_{\phi} S/\Delta S}$, where $\partial_{\phi} \equiv \partial/\partial \phi$. Here, $\partial_{\phi} S$ is the phase gradient of the signal (PGS), and $\Delta S$ is the standard deviation of the signal (SDS). When excess noise (EN) is suppressed sufficiently, $\Lambda$ is limited by the quantum projection noise (QPN)~\cite{Itano}, and is given by the inverse of the quantum phase fluctuation (QPF$^{-1}$). For  a conventional atomic interferometer, the sensitivity is at the Standard Quantum Limit (SQL): $\Lambda=\text{QPF}^{-1}= \sqrt{N}$, with $N$ being the number of atoms interrogated within the measurement time. Using spin-squeezing, it is possible to surpass the SQL, and a key goal in this context is to achieve the Heisenberg Limit (HL), under which $\Lambda=N$, representing an improvement by a factor of $\sqrt{N}$.

To enhance  $\Lambda$, one can either increase the PGS or decrease the SDS. In a conventional approach for spin squeezing, one minimizes the SDS. For example, using optimal one-axis-twist squeezing (OATS) and two-axis-counter-twist (TACT) squeezing ~\cite{Kitagawa}, the SDS can be reduced respectively by a factor of $N^{1/3}$ and $\sqrt{N/2}$, while the PGS remains essentially unchanged, compared to those of a conventional atomic interferometer. As such, $\Lambda=N^{5/6}$ for the former and $\Lambda=N/\sqrt{2}$ for the latter. Though the TACT squeezing can yield a better sensitivity, it is experimentally more complicated than the OATS~\cite{Helmerson,Bouchoule,Zhang1,Schleier,Leroux,Zhang2,Foss,Hosten1}. Recently~\cite{Davis,Hosten2,Sarkar1}, it was shown that it is also possible to reach sensitivity at or near the HL using variants of the OATS. Ref. ~\cite{Davis} proposed and Ref. ~\cite{Hosten2} demonstrated the echo squeezing protocol (ESP), which can increase the PGS by a factor of $\sim \sqrt{N/e}$, while leaving the SDS unchanged, thus producing $\Lambda \approx N/\sqrt{e}$. In ref.~\cite{Sarkar1} we proposed a Schr\"odinger Cat atomic interferometer (SCAIN) that makes use of critically tuned OATS, rotation, inverse rotation and unsqueezing, which, in combination with collective state detection (CSD)~\cite{Dicke,Arecchi,Sarkar2,Kim,Sarkar3}, reduces the SDS by a factor of $\sqrt{N}$, while leaving the PGS unchanged, yielding $\Lambda=N$. In what follows, we will refer to this as the CSD-SCAIN.

In this paper, we describe a new protocol that is a variant of the CSD-SCAIN protocol, with radically different behavior.  It employs the conventional detection (CD) technique by measuring directly the populations of the spin-up or spin-down states of individual atoms.  We show that, under this protocol (called CD-SCAIN), the PGS is increased by a factor of $N$, while the SDS is also \textit{increased} by a factor of $\sqrt{N}$.  The net enhancement in $\Lambda$ is by a factor of $\sqrt{N}$, reaching the HL. However, because of the increase in noise (i.e., SDS), this is now significantly more robust to excess noise (EN) than all the protocols described above.  \textit{Specifically, for this protocol, it should be possible to achieve} $\Lambda=N/\sqrt{2}$ \textit{even when the EN is greater than the QPN for a conventional atomic interferometer by a factor of} $\sqrt{N}$.

\begin{figure}[h]
\includegraphics[scale=0.34]{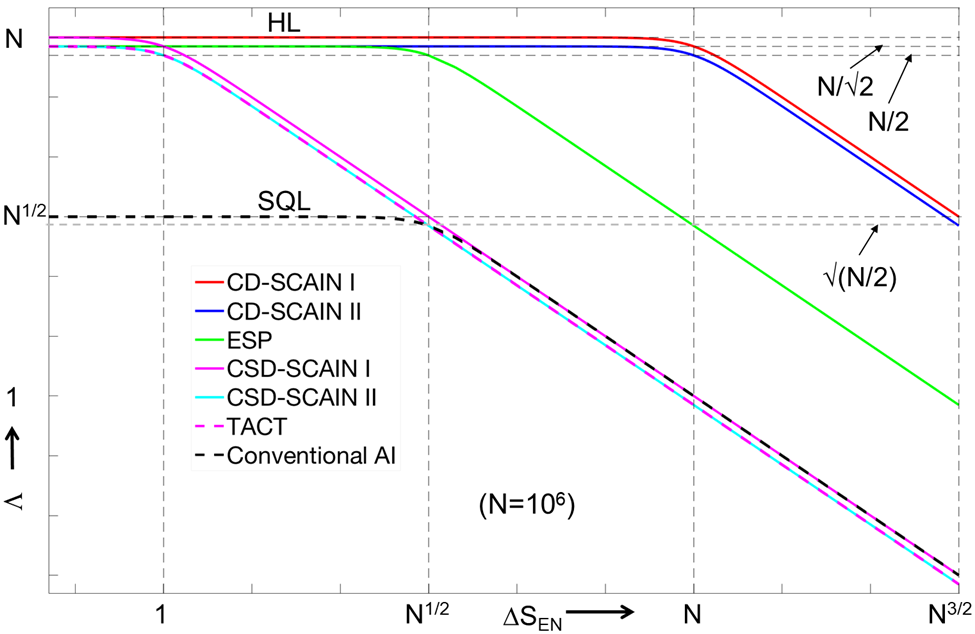}
\caption{The sensitivity, $\Lambda$, as a function of excess noise, $\Delta S_{\text{EN}}$, for various protocols.  For both CSD-SCAIN and CD-SCAIN, we have used two labels: I and II; I indicates the case when the parity of N is known, while II indicates the case where the signal is averaged over both parities.}
\label{fig:1}
\end{figure}

The degree of suppression of EN for different protocols is illustrated in Fig.~\ref{fig:1}.  Here, we consider a situation where EN contributes an additional variance, $\Delta S^2_{\text{EN}}$, to the signal.  The sensitivity is then given by $\Lambda=\abs{\text{PGS}/\sqrt{\Delta S^2_{\text{QPN}}+\Delta S^2_{\text{EN}}}}=\Lambda_{\text{QPN}}/\sqrt{1+\rho^2}$, where $\rho \equiv \Delta S_{\text{EN}}/\Delta S_{\text{QPN}}$.  One way to characterize the degree of robustness against EN is by determining the value of $\Delta S_{\text{EN}}$ for which $\rho=1$.  As can be seen,  for TACT, this value is 1, making it particularly vulnerable to EN.  In contrast, for ESP (as well as for the conventional atomic interferometer), this value is $\sqrt{N}$, making it a factor of $\sqrt{N}$ more robust than TACT.  For CD-SCAIN, this value is $N$, making it a factor of $\sqrt{N}$ ($N$) more robust than ESP (TACT).  We also see that CSD-SCAIN is as sensitive to EN as TACT.  Thus, in switching from collective state detection to conventional detection, the robustness of the SCAIN protocol to EN is improved by a factor of $N$.  One can also define the range of usefulness of a protocol as the value of $\Delta S_{\text{EN}}$ for which the sensitivity drops to $\Lambda=\sqrt{N/2}$.  By this measure, the usefulness of CD-SCAIN  extends to $N^{3/2}$, while that for ESP extends only to $N$.  A systematic study of the robustness of various spin-squeezin protocol has been carried out in Ref. ~\cite{Haine3}.  This study cites an earlier version of this paper posted on the arxiv by us (~\cite{Fang}) as having the greatest robustness against excess noise.

The rest of the paper is organized as follows.  In Section ~\ref{Section 2} we describe the protocol for the CD-SCAIN: the Schr\"odinger Cat Atomic Interferometer using Conventional Detection.   In Section ~\ref{Section 3}, we present the analytical model for the CD-SCAIN, and discuss additional details underlying the CD-SCAIN protocol.  In Section ~\ref{Section 4} we describe the protocol for the CD-SCAC: the Schr\"odinger Cat Atomic Clock using Conventional Detection.   In Section ~\ref{Section 5} we discuss experimental considerations, including a detailed analysis of the one axis twist scheme based on the cavity feedback approach, and estimate the effect of cavity decay and spontaneous emission on the fidelity of the CD-SCAIN and the CD-SCAC.  In Section ~\ref{Section 6}, we discuss comparisons with other proptocols (~\cite{Haine1, Haine2, Huang}) that are closely related to ours.  Finally, we present the conclusion in  Section ~\ref{Section 7}.

\section{Schr\"odinger Cat Atomic Interferometer Using Conventional Detection}
\label{Section 2}

\begin{figure}[h]
\includegraphics[scale=0.278]{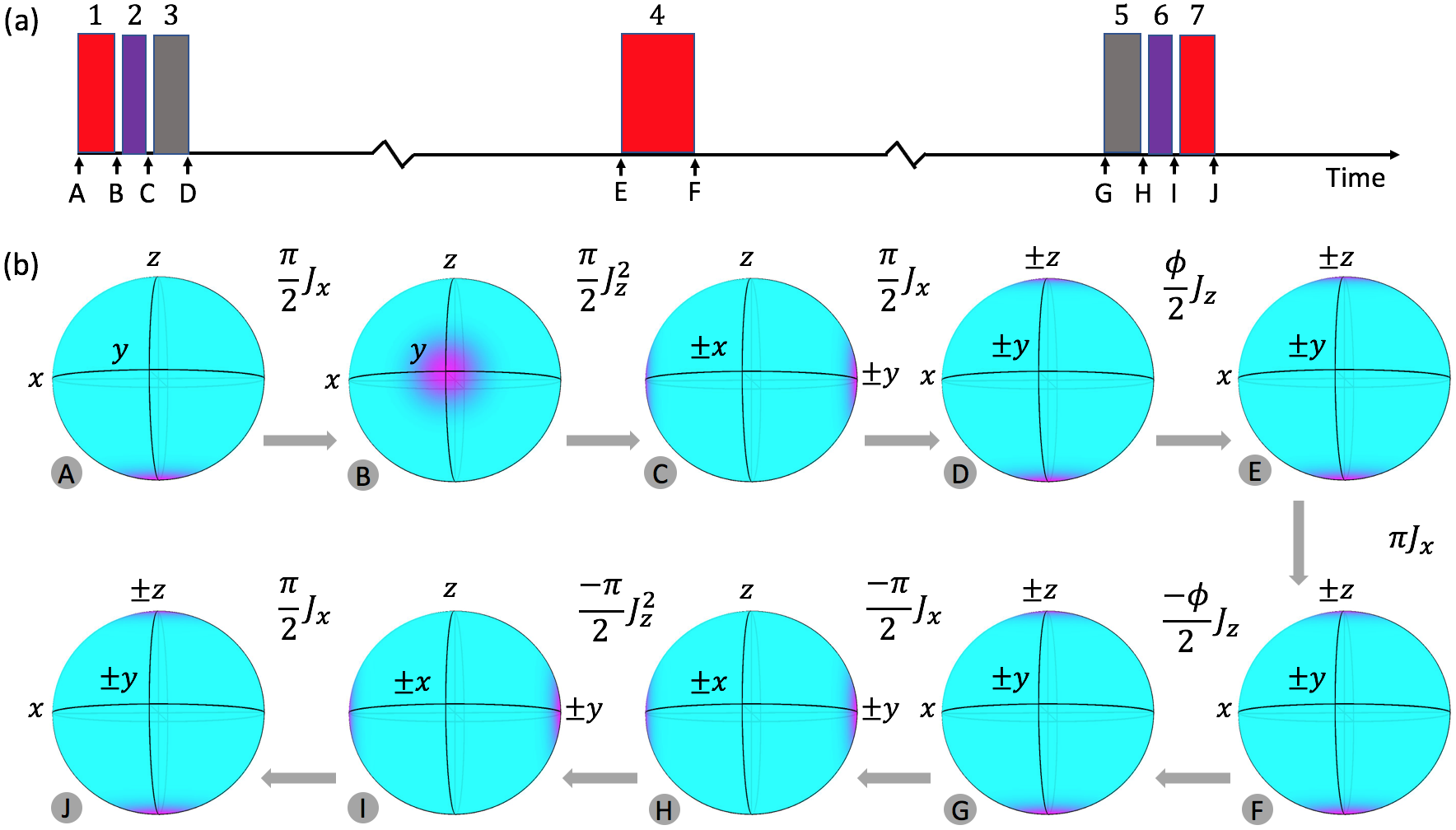}
\caption{(a) Schematic illustration of the protocol employed for the Schr\"odinger Cat Atomic Interferometer (SCAIN). (b) The Husimi Quasi Probability Distributions (QPDs) at different stages of the protocol, for $N=40$, $\mu = \pi/2$, $\text{ARA}=\vu{x}$, $\xi = -1$ and $\phi = 0.5\pi/N$.}
\label{fig:2}
\end{figure}

The atomic interferometer considered here is a SCAIN, which is based on the conventional Raman atomic interferometer (CRAIN)~\cite{Borde,Kasevich,Riehle,Shahriar1}. Briefly, both make use of $N$ three-level atoms with metastable states $\ket{1, p_z = 0}$ and $\ket{2, p_z = \hbar k}$ and an excited state $\ket{3}$ in the $\Lambda$-configuration, coupled by a pair of counter-propagating laser beams. Here, $k \equiv k_1 + k_2$, with $k_1$ ($k_2$) being the wave number for the beam propagating in the $+\vu{z}$ ($-\vu{z}$) direction, and $p_z$ is the $z$-component of the linear momentum of the atom. Each atom can be reduced to an equivalent two-level model via adiabatic elimination of the excited state~\cite{Shahriar2,Shahriar3}, and thus can be represented by a pseudospin-$1/2$ operator $\vu{j}$, where we define $\ket{\downarrow} \equiv \ket{1, p_z = 0}$ and $\ket{\uparrow} \equiv \ket{2, p_z = \hbar k}$. The ensemble, now represented by a collective spin operator $\vu{J} \equiv \sum_{i}^{N}\vu{j}_i$, is initially prepared in a coherent spin state ~\cite{Arecchi}, $\ket{-\vu{z}} = \prod_{i=1}^{N}\ket{\downarrow}$, where all atoms are in the spin-down state. Here we employ the notation that a state $\ket{\vu{e}}$ is a coherent spin state in the direction of the unit vector $\vu{e}$, with the pseudospin vector of each atom being in that direction. For the CRAIN, the ensemble is then subjected to the usual pulse sequence of $\pi/2-$dark$-\pi-$dark$-\pi/2$, labeled as $1$, $4$, $7$ in Fig.~\ref{fig:2}~(a). For the SCAIN, however, the ensemble will undergo four additional pulses labeled as $2$, $3$, $5$, $6$ in Fig.~\ref{fig:2}~(a), corresponding to the squeezing, rotation, inverse rotation and unsqueezing operations, as described in the CSD-SCAIN protocol proposed in Ref.~\cite{Sarkar1}.

The complete evolution of the quantum states on a Bloch sphere under this protocol is shown in Fig.~\ref{fig:2}~(b), using the Husimi Quasi Probability Distribution (QPD)~\cite{Kitagawa,Arecchi}. It should be noted that the exact effects of the protocol depend on the choices of a set of parameters such as the value (and parity) of $N$, the squeezing parameter $\mu$ for the OATS, the auxiliary rotation axis (ARA, can be $\vu{x}$ or $\vu{y}$ axes) around which to implement the rotation, the corrective rotation sign $\xi$ which can take values of $\pm1$ corresponding to redoing or undoing the first auxiliary rotation, and lastly the dark zone phase shift $\phi$. The case shown here is for an even value of $N=40$, with $\mu=\pi/2$, $\text{ARA}=\vu{x}$, $\xi=-1$ and $\phi=\pi/80$. The QPD is expressed as a function $Q_H(\theta, \phi)$ of the angles in spherical coordinates which span the surface of the Bloch sphere. For a given quantum state $\ket{\Psi}$, it is defined as $Q_H(\theta, \phi) \equiv \abs{\ip{\Psi}{\Phi(\theta, \phi)}}^2$, where
\begin{equation}
\ket{\Phi(\theta, \phi)} \equiv \qty(\cos\frac{\theta}{2})^N \sum_{k=0}^{N}\sqrt{\binom{N}{k}}\qty(e^{i\phi}\tan{\frac{\theta}{2}})^k \ket{E_{N-k}}
\label{eq:1}
\end{equation}
represents the coherent spin state corresponding to all the spins pointing in the direction $\{\theta, \phi \}$, and $\ket{E_n}$ are the Dicke collective states   ~\cite{Dicke,Arecchi,Sarkar2} defined as
\begin{equation}
\ket{E_n} = \sum_{k=1}^{\binom{N}{n}}P_k\ket{\downarrow^{N-n}\bigotimes\uparrow^{n}}\bigg{/}\sqrt{\binom{N}{n}}
\label{eq:2}
\end{equation}
with $P_k$ being the permutation operator~\cite{Hume}. Here, the extremal state $\ket{E_N}$ corresponds to all pseudospins in the $\vu{z}$ direction. As such, we will refer to these as the $Z$-directed Dicke Collective States (ZDCSs). As needed, we will also refer to XDCSs (YDCSs) for which $\ket{E_N}$ corresponds to all pseudospins in the $\vu{x}$ ($\vu{y}$) direction.

In illustrating the nature of the QPD at various stages, we have used different orientations of the Bloch sphere as suited, and added $\pm$ symbols in front of two axes to indicate that the picture looks the same when it is rotated by $180$ degrees around the third axis. At the start (point A), the system is in state $\ket{-\vu{z}}$. After the first $\pi/2$ pulse (point B), the state rotates around the $\vu{x}$ axis to reach state $\ket{\vu{y}}$. We then apply a squeezing Hamiltonian of the form $H_{OATS} = \chi \hat{J}_z^2$ for a duration of $\tau$ such that $\mu=\chi\tau$. After the squeezing pulse (point C), the state is split equally between two coherent spin states, and can be expressed as $(\ket{\vu{y}} - \eta\ket{-\vu{y}})/\sqrt{2}$~\cite{Sorensen,Leibfried1,Leibfried2,Leibfried3,Blatt}, where $\eta = i(-1)^{N/2}$, representing a phase factor with unity amplitude.  It should be noted that this phase factor depends on the super-even-parity, representing whether $N/2$ is even or odd; however, the shapes of the fringes, as well as the values of $\text{QFR}^{-1}$, for both CSD and CD protocols, are not expected to depend on the value of the super-even-parity, as we have verified explicitly. 

This is a Schr\"odinger Cat (SC) state~\cite{Schroedinger}, but as a superposition of the two extremal states of the YDCS manifold, which cannot be used to achieve phase magnification, since the phase difference between the two arms corresponds to rotation around the $\vu{z}$ axis. This problem is solved by applying the auxiliary rotation of $\pi/2$ around the $\vu{x}$ axis, which transforms this state to $(\ket{-\vu{z}} + \eta\ket{\vu{z}})/\sqrt{2}$. This (point D) represents the desired SC state, as a superposition of the two extremal states of the ZDCS manifold: $(\ket{E_0} + \eta\ket{E_N})/\sqrt{2}$. After the first dark zone (point E), the state is $e^{-i\phi\hat{J}_z/2} \qty(\ket{E_0}_L + \eta\ket{E_N}_U)/\sqrt{2}$, where the subscript $L$ ($U$) is for the lower (upper) arm of the interferometer (the total phase shift $\phi$ is split equally in the two dark zones, as discussed in Section ~\ref{Section 3}, Subsection H). Since both $\ket{E_0}$ and $\ket{E_N}$ are eigenstates of $\hat{J}_z$, with eigenvalues of $-N/2$ and $N/2$ respectively ($\hbar = 1$), this state can be simplified to $\qty(e^{i\phi N/4} \ket{E_0}_L + e^{-i\phi N/4} \eta\ket{E_N}_U)/\sqrt{2}$. The resulting QPD remains unchanged but the quantum state incorporates these phase accumulations. After the $\pi$-pulse (point F), $\ket{E_0}_L$ becomes $-i\ket{E_N}_L$ while $\ket{E_N}_U$ becomes $-i\ket{E_0}_U$. After the second dark zone (point G), the state is $\qty(e^{i\phi N/2} \eta\ket{E_N}_L + e^{-i\phi N/2} \ket{E_0}_U)/\sqrt{2}$, so that the net phase difference between the two paths is $N\phi$, thus magnifying the rotation induced phase by a factor of $N$. To reveal the phase magnification, we apply another auxiliary rotation by an angle of $-\pi/2$ around the $\vu{x}$ axis (point H), followed by the unsqueezing Hamiltonian $-H_{OATS}$ (point I). After the second $\pi/2$ pulse (point J) the state is $\ket{\Psi}_f = \cos(N\phi/2)\ket{E_0} - \eta\sin(N\phi/2)\ket{E_N}$. The whole protocol can be expressed as:
\begin{multline}
\ket{\Psi}_f = e^{-i\frac{\pi}{2}\hat{J}_x} e^{i\mu\hat{J}_z^2} e^{-i\xi\frac{\pi}{2}\hat{J}_x} e^{i\frac{\phi}{2}\hat{J}_z} e^{-i\pi\hat{J}_x} \\
e^{-i\frac{\phi}{2}\hat{J}_z} e^{-i\frac{\pi}{2}\hat{J}_x} e^{-i\mu\hat{J}_z^2} e^{-i\frac{\pi}{2}\hat{J}_x} \ket{-\vu{z}}
\label{eq:3}
\end{multline}

If the population of the collective state $\ket{E_0}$ were detected, the signal would be $\cos[2](N\phi/2)$, with fringes a factor of $N$ narrower than that for the CRAIN; this is the CSD-SCAIN ~\cite{Sarkar1}. Compared to the CRAIN, the phase gradient of the signal (PGS) remains unchanged, since the phase enhancement is countered by reduction in the signal amplitude by a factor of $N$. However, the standard deviation of the signal (SDS) is now reduced by a factor of $\sqrt{N}$, since the number of particles is unity.  As such, the sensitivity increases by $\sqrt{N}$, reaching the HL.  In what follows, we describe a significantly different version of the SCAIN, namely the CD-SCAIN, which employs the conventional detection technique corresponding to measuring the $z$-component of the combined spin of all atoms, the $\hat{J_z}$ operator, which represents the difference between the number of atoms in the spin-up and spin-down states. 

\begin{figure}[h]
\includegraphics[scale=0.45]{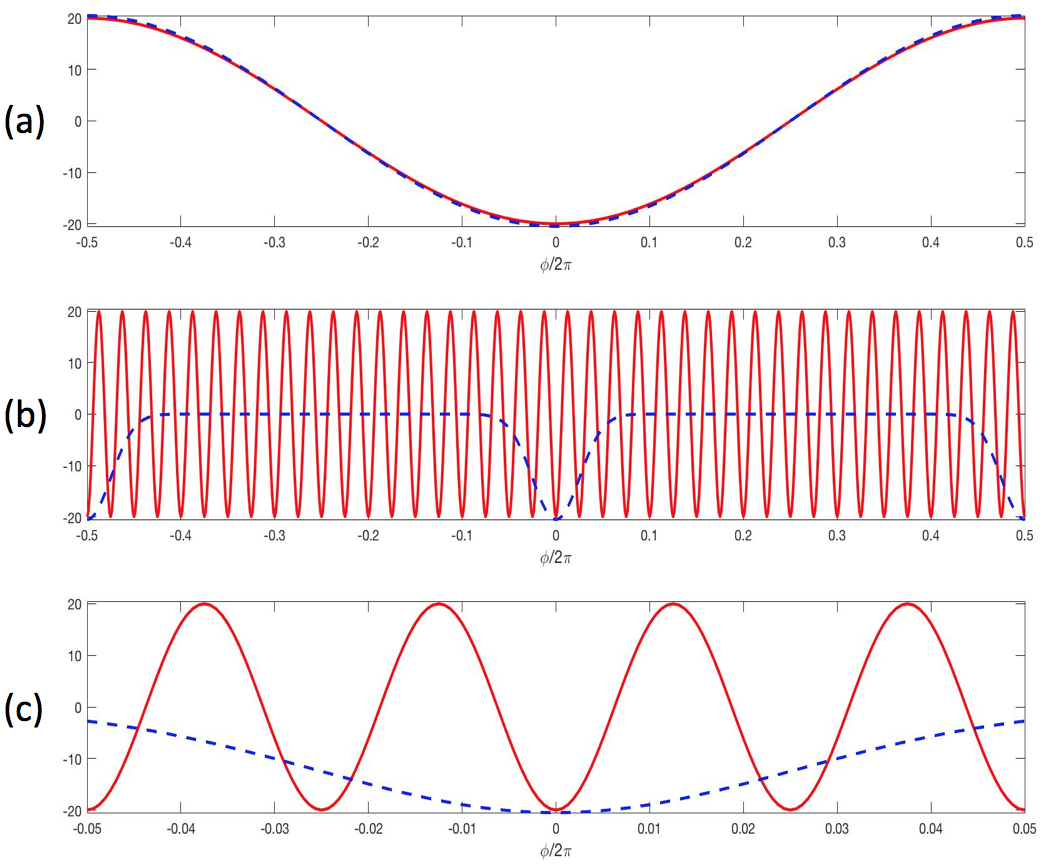}
\caption{Signals corresponding to detection of $\ev*{\hat{J}_z/\hbar}$, as a function of $\phi$, for $\mu=\pi/2$, $\text{ARA}=\vu{x}$ and $\xi = -1$. $N=40$ is red while $N=41$ is dashed-blue. (a) Fringes for CRAIN for comparison; (b) Fringes for CD-SCAIN; (c) Zoomed-in fringes for CD-SCAIN. The horizontal span in (c) is $10$ times smaller than those in (a) and (b).}
\label{fig:3}
\end{figure}

The signal for the CD-SCAIN is obtained by expanding  $\hat{J_z}$ in the basis of the ZDCSs, then taking the expectation value with respect to $\ket{\Psi}_f$. This is found to be $\ev{\hat J_z}{\Psi_f}=-N/2\cos(N\phi)$, as derived in Section ~\ref{Section 3}, again showing $N$-fold fringe narrowing.  However, compared to the case of the CSD-SCAIN, the amplitude of the fringes is now a factor of $N$ larger. As such, \textit{the phase gradient of the signal (PGS) is now larger than that for a CRAIN by a factor of N} .  At the same time, \textit{the standard deviation of the signal (SDS) is also increased by a factor of} $\sqrt{N}$, compared to that for a CRAIN, as derived and discussed further in Section ~\ref{Section 3}.  This is surprising, since the signal amplitude for the CD-SCAIN is the same as that for a CRAIN.  The net enhancement in sensitivity is by a factor of $\sqrt{N}$, reaching the HL, just as in the case of the CSD-SCAIN.  However, the increase in SDS makes the CD-SCAIN significantly more robust against excess noise (EN), as summarized earlier in Fig.~\ref{fig:1}.

For the particular choice of the auxiliary rotation axis (ARA) used in the protocol for Fig.~\ref{fig:2}~(b), the expression for the signal for the CD-SCAIN shown above applies only to the case when $N$ is even. The results for odd value of $N=41$ with all other parameters  the same as in Fig.~\ref{fig:2}~(b) are found to be drastically different (see Section ~\ref{Section 3}), due to the fact that the state after the squeezing pulse will now be split equally between $\ket{\vu{x}}$ and $\ket{-\vu{x}}$, thus generating an SC state as a superposition of the two extremal states of the XDCS manifold~\cite{Sorensen,Leibfried1,Leibfried2}.  This modification of the state, caused by a change of just 1 in the value of N, can be understood by noting that the propagator corresponding to the OATS Hamiltonian for $\mu=\pi/2$ can be decomposed as a sum of two parts, one of which is proportional to a product of N Pauli-spinors (as shown in ref. \cite{Leibfried1} and proven analytically in ref. \cite{Leibfried2}).  The ensuing auxiliary rotation around the $x$ axis will not transform it into the desired SC state required to yield the $N$-fold phase amplification. This also complicates the evolution of the quantum states during the following stages, for which an analytical expression for the final state is not easy to find. Instead, we take a numerical approach to simulate the state evolutions, as discussed in Section ~\ref{Section 3}. The signals for the CD-SCAIN, as a function of $\phi$, for both even and odd values of $N$, are shown in Fig.~\ref{fig:3}, where for reference, the signal corresponding to one full fringe of the CRAIN is also shown in Fig.~\ref{fig:3}~(a). The plots in Fig.~\ref{fig:3}~(b) and~(c) clearly show the $N$-fold narrowed fringes for the even case while only a central fringe is observable for the odd case. We also find that changing the sign of $\xi$ simply inverts the fringes, which implies that the $N$-fold reduction of the fringe width happens for the even case no matter whether we choose to redo ($\xi=1$) or undo ($\xi=-1$) the first auxiliary rotation.  Of course, the nature of the signals for odd and even values of $N$ can be reversed by switching the choice of the auxiliary rotation axis (ARA) from $\vu{x}$ to $\vu{y}$.

\begin{figure}[h]
\includegraphics[scale=0.26]{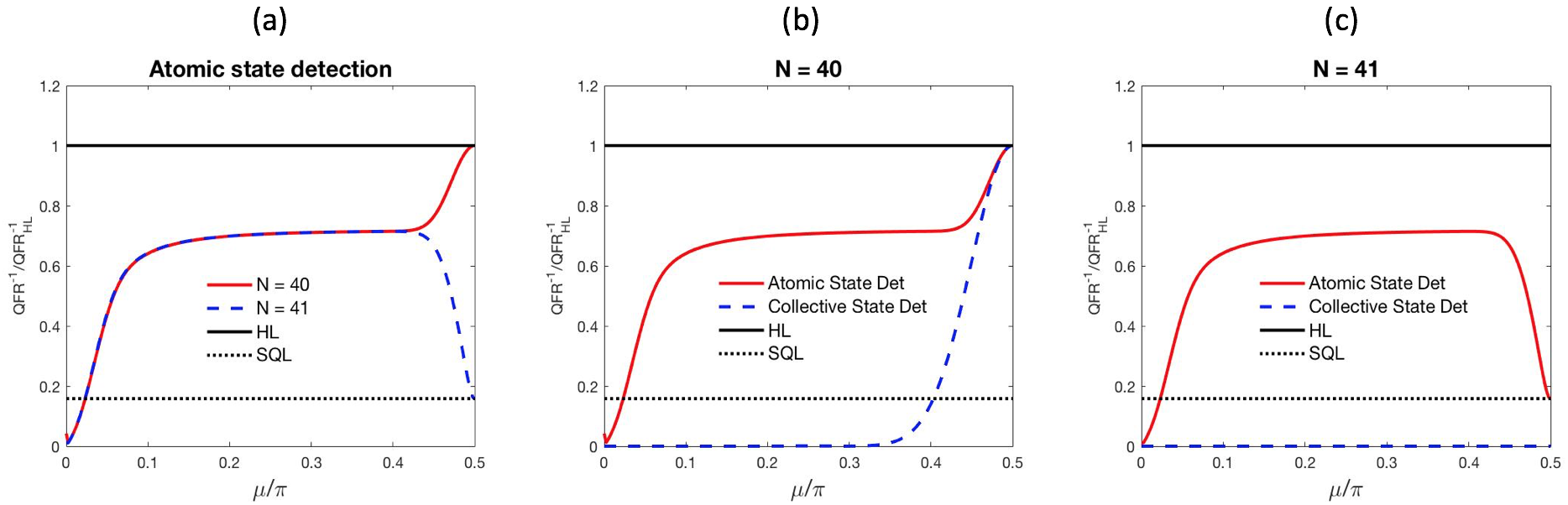}
\caption{Illustration of $\text{QFR}^{-1}$ for different cases, as a function of the squeezing parameter $\mu$, normalized to the HL (solid black line), for $\text{ARA}=\vu{x}$ and $\xi=+1$. (a) The case for the CD-SCAIN, with red for $N=40$ and dashed-blue for $N=41$; (b) Comparison between the CD-SCAIN and the CSD-SCAIN for even $N=40$; (c) Comparison between the CD-SCAIN and the CSD-SCAIN for odd $N=41$. The dotted black line shows the SQL.}
\label{fig:4}
\end{figure}

In Fig.~\ref{fig:4}, we illustrate the behavior of the inverse of the quantum fluctuation in rotation (QFR$^{-1}$), as a function of the squeezing parameter $\mu$, for different choices of parameters for the CD-SCAIN, along with a comparison with the CSD-SCAIN. The QFR$^{-1}$ is a special case of QPF$^{-1}$ when the the phase difference is induced by rotation. For each case, the QFR$^{-1}$ is normalized to the QFR$^{-1}_{\text{HL}}$ for $N=40$, indicated as the solid black line. The dashed black line shows the QFR$^{-1}_{\text{SQL}}$ for $N=40$. Fig.~\ref{fig:4}~(a) shows the QFR$^{-1}$ for the CD-SCAIN only. For $\mu=\pi/2$, the sensitivity for even number of atoms (red) is at the HL, and that for odd number of atoms (dashed blue) is at the SQL. For even $N$, this sensitivity is reached due to an amplification of phase by a factor of $N$, and a concomitant increase in the standard deviation of the signal (SDS) by a factor of $\sqrt{N}$. For odd $N$, there is a phase amplification, manifested as a Fabry-Perot like fringe around $\phi=0$ which is narrowed by a factor of $\sqrt{N}$, along with an increase in the standard deviation of the signal (SDS) by a factor of $\sqrt{N}$. The difference between the two cases disappears when the value of $\mu$ is reduced below a threshold value of $\sim0.45\pi$. There is a range of values of the squeezing parameter ($0.2\pi\le\mu\le0.45\pi$) over which the normalized value of QFR$^{-1}$ is $\sim1/\sqrt{2}$.  Finally, we note that the vanishing value of QFR$^{-1}$ for $\mu=0$ is simply due the fact that the signal is constant as a function of $\phi$. In Fig.~\ref{fig:4}~(b) and~(c), we compare the sensitivity of the CD-SCAIN with that of the CSD-SCAIN, for even and odd $N$, respectively. For even $N$, the sensitivity for both detection protocols are the same for $\mu=\pi/2$. However, for the CSD-SCAIN, the sensitivity drops off to zero rapidly for decreasing values of $\mu$. For odd values of $N$, the sensitivity for the CSD-SCAIN is zero for all values of $\mu$, due to the signal being a constant as a function of $\phi$. For both odd and even values of $N$, the results for the CD-SCAIN are the same for both values of $\xi$($=\pm1$), while the results for CSD-SCAIN shown here is for $\xi=+1$. The CSD-SCAIN result for $\xi=-1$ is qualitatively the same, with slight differrences for small values of $\mu$.

Until now, we have analyzed and compared the performance of CD-SCAIN in a separate manner for even and odd values of $N$.   In scenarios where the odd and even parity cases can occur with equal probablities (for example, when atoms  caught in a magneto-optic trap and released are used as the source for the CD-SCAIN), the average value of $\text{QFR}^{-1}$ is given by: $\text{QFR}^{-1}_{\text{AVE}} = \qty[(\text{QFR}^{-1}_{\text{EVEN}})^2/2 + (\text{QFR}^{-1}_{\text{ODD}})^2/2]^{1/2}$, as derived in Section ~\ref{Section 3}, subsection G.  Thus, for a large number of atoms ($N\gg1$), the average sensitivity is a factor of $\sqrt{2}$ below the HL.

Very similar results can be obtained for an atomic clock as well. The behavior of a Schr\"odinger Cat Atomic Clock (SCAC) under conventional detection (CD-SCAC) and its comparison with a SCAC under collective state detection (CSD-SCAC) are presented in  Section ~\ref{Section 4}.

\section{Analytical Model and Additional Details for the CD-SCAIN}
\label{Section 3}

In this section, we provide some additional details for understanding the SCAIN employing the conventional detection (CD) protocol, and its comparisons with the SCAIN employing the collective state detection (CSD) protocol.

\subsection{Matrix elements of the collective spin operators}

As discussed earlier, the squeezing pulse complicates the evolution of the quantum states for the ensemble and it is generally not easy to write down explicitly the mathematical expressions for the final states for  arbitrary values of $\phi$ (the phase difference) and $\xi$ (the corrective rotation sign). Therefore a numerical approach is employed to simulate the evolutions for each stage of the protocol. The basis of the operators are chosen to be the Dicke collective states defined earlier in ~\ref{eq:2}, which are the eigenstates of the $\hat{J}_z$ operator, with eigenvalues ranging from $-N\hbar/2$ for the $\ket{E_0}$ state to $N\hbar/2$ for the $\ket{E_N}$ state.  In general, for a total spin of $J=N/2$, the eigenstate corresponding to an eigenvalue of $m\hbar$ will be $\ket{E_{J+m}}$. The matrix elements of the relevant operators can thus be expressed as follows ~\cite{Dicke}:
\begin{align}
& \mel{E_{J+m'}}{\hat{J}_x}{E_{J+m}} = \frac{\hbar}{2}(A_{J,m}\delta_{m',m+1} + B_{J,m}\delta_{m',m-1}) \nonumber \\
& \mel{E_{J+m'}}{\hat{J}_y}{E_{J+m}} = \frac{\hbar}{2i}(A_{J,m}\delta_{m',m+1} - B_{J,m}\delta_{m',m-1}) \nonumber \\
& \mel{E_{J+m'}}{\hat{J}_z}{E_{J+m}} = \hbar m \delta_{m',m} \nonumber \\
& \mel{E_{J+m'}}{\hat{E}_{n',n}}{E_{J+m}} = \delta_{J+m',n'}\delta_{J+m,n}
\label{eq:5}
\end{align}
where $\hat{E}_{n',n} \equiv \dyad{E_{n'}}{E_n}$ is the projection operator for the collective states, and $A_{J,m} = \sqrt{(J-m)(J+m+1)}$ and $B_{J,m} = \sqrt{(J+m)(J-m+1)}$ are the two normalization coefficients associated with the raising and lowering operators, respectively. For all the results shown in the main text and the supplements, we have made use of these $(N+1)\times(N+1)$ matrices to represent all operators, and carried out the complex matrix exponentiations numerically.

\subsection{Derivation of $\text{QFR}^{-1}$ for the CSD-SCAIN and CD-SCAIN protocols}
As shown earlier, with the chosen parameters, the final state of the ensemble for both the CSD-SCAIN and CD-SCAIN protocols is given by $\ket{\Psi}_f = \cos(N\phi/2)\ket{E_0} - \eta\sin(N\phi/2)\ket{E_N}$. For the CSD-SCAIN protocol, in general the collective state operator to be measured can be defined as $\hat{Q}_{M,CSD,m} \equiv \op{E_m}$. Thus, the operator we measure is $\hat{Q}_{M,CSD,0}$ if we detect the $\ket{E_0}$ state, and $\hat{Q}_{M,CSD,N}$ if we detect the $\ket{E_N}$ state. For the final state described above, if we measure the former, the signal is $\cos[2](N\phi/2)$; if we measure the latter, the signal is $\sin[2](N\phi/2)$. For the CD-SCAIN protocol, the operator we measure is $\hat{Q}_{M,CD} = \hat{J}_z/\hbar$. From the third line of Eq.~\ref{eq:5}, it follows that $\hat{Q}_{M,CD} = \hat{J}_z/\hbar = \sum_{m=-J}^{J}m\op{E_{J+m}} = \sum_{m=-J}^{J}m\hat{Q}_{M,CSD,J+m}$. In the final state described above, we have only two of the collective states. As such, for this state, $\ev*{\hat{Q}_{M,CD}} = -J\ev*{\hat{Q}_{M,CSD,0}} + J\ev*{\hat{Q}_{M,CSD,N}}$. Thus it follows that for the CD-SCAIN protocol, the signal is given by $-J\cos[2](N\phi/2) + J\sin[2](N\phi/2) = -(N/2)\cos(N\phi)$, which has the same fringe width as that obtained by using the CSD protocol, except that the signal now ranges from $N/2$ to $-N/2$.

To determine the $\text{QFR}^{-1}$ for both protocols, we define first the signal for the CSD-SCAIN as $\Sigma \equiv \ev*{\hat{Q}_{M,CSD,0}} = \cos[2](N\phi/2)$ and the standard derivation of the signal (SDS) as $\Delta\Sigma \equiv [\ev*{\hat{Q}^2_{M,CSD,0}} - \Sigma^2]^{1/2}$. Similarly we define the signal for the CD-SCAIN as $S \equiv \ev*{\hat{Q}_{M,CD}} = -(N/2)\cos(N\phi)$ and the standard deviation of the signal (SDS) as $\Delta S \equiv [\ev*{\hat{Q}^2_{M,CD}} - S^2]^{1/2}$. Noting that $\phi = 2mA\Omega_G/\hbar \equiv \Omega_G/\Gamma$, with $A$ being the area of the whole interferometer and $\Omega_G$ being the normal component of the rate of rotation, we can now write:
\begin{align}
& \text{QFR}^{-1}_{CSD-SCAIN} = \abs{\Gamma^{-1} \frac{\pdv*{\Sigma}{\phi}}{\Delta\Sigma}} \nonumber \\
& \text{QFR}^{-1}_{CD-SCAIN} = \abs{\Gamma^{-1} \frac{\pdv*{S}{\phi}}{\Delta S}}
\label{eq:6}
\end{align}

For the CSD-SCAIN protocol, we note that $\hat{Q}^2_{M,CSD,0} = \hat{Q}_{M,CSD,0}$. which means that $\Delta\Sigma \equiv [\Sigma - \Sigma^2]^{1/2}$. Using the expression for $\Sigma$ from above, we easily find that QFR$^{-1}_{CSD-SCAIN} = N/\Gamma$. We recall that the value of QFR$^{-1}$ for a CRAIN is given by QFR$^{-1}_{CRAIN} = \sqrt{N}/\Gamma$, which is the SQL. As such, the CSD-SCAIN represents an improvement by a factor of $\sqrt{N}$, reaching the HL sensitivity. 

For the CD-SCAIN protocol, we see that $\hat{Q}^2_{CD} = \sum_{m=-J}^{J}m^2\op{E_{J+m}}$. However, in the final state described above, we have only two of the collective states. As such, we get $\ev*{\hat{Q}^2_{CD}} = J^2\ev*{\hat{Q}_{M,CSD,0}} + J^2\ev*{\hat{Q}_{M,CSD,N}} = J^2 = N^2/4$. Thus, it follows immediately that $\Delta S \equiv \qty[\ev*{\hat{Q}^2_{CD}} - S^2]^{1/2} = \qty{N^2/4 - N^2/4\qty[\cos[2](N\phi)]}^{1/2} = (N/2)\abs{\sin(N\phi)}$. It should be noted that the peak value of the standard deviation of the signal (SDS) in this case is $N/2$, which happens at the points where the slope of the fringe is maximum. From the second line of Eq.~\ref{eq:6}, we then get QFR$^{-1}_{CD-SCAIN} = N/\Gamma$, the same as that for the CSD-SCAIN protocol, yielding the HL sensitivity.

In the context of atomic interferometers, one often makes use of a rule-of-thumb that states that the quantum fluctuation in rotation (QFR) is simply given by the linewidth (as a function of rotation rate) divided by the signal to noise ratio (SNR), being equal to the square-root of the number of particles. For the case of the CRAIN, the $\text{QFR}$ is given by $\Gamma/\sqrt{N}$, where $\Gamma$ is the linewidth (representing an amount of rotation that produces a phase shift of one radian) and $\sqrt{N}$ is the SNR, so the above rule-of-thumb applies. For the case of the CSD-SCAIN, the linewidth is reduced by a factor of $N$ compared to that of the CRAIN. But the SNR is also reduced by a factor of $\sqrt{N}$, since the number of particles is now unity, not $N$. Thus, according to this rule-of-thumb, the QFR of the CSD-SCAIN should be $\Gamma/N$. This is consistent with what is found above for this case. For the case of the CD-SCAIN, however, if we try to apply the same rule-of-thumb, we reach an erroneous conclusion. While the linewidth for the CD-SCAIN is also reduced by a factor of $N$, there is no reduction in the number of particles, since the fringe amplitude is $N$, the same as that for the CRAIN. This in turn would imply that the SNR remains the same, so the QFR would be $\Gamma/N^{3/2}$, thus exceeding the HL by a factor of $\sqrt{N}$. This suggests that \textit{the above rule-of-thumb is not applicable to the case of the CD-SCAIN}, where, in fact, the SNR is also reduced by a factor of $\sqrt{N}$ instead of staying unchanged, due to the nature of the SC state, as shown above.

\subsection{Distinction between the CD-SCAIN and CSD-SCAIN protocols for general quantum states}

In this subsection, we show mathematically the distinction between the CD-SCAIN and CSD-SCAIN protocols for general quantum states. Let us define as $\hat{q}_M$ the operator for each atom whose expectation value is measured during the experiment. For each atom, let us define $\ket{e}$ ($\ket{g}$) to be the spin-up (-down) state. Thus, we can write $\hat{q}_M = \mu_g\op{g} + \mu_e\op{e}$, where $\mu_g$ and $\mu_e$ are complex numbers. The operator which is measured for the whole system can be expressed as $\hat{Q}_M = \sum_{k=1}^{N}\hat{q}_{M,k}$. We can express the quantum state of each atom as $\ket{\psi} = C_g\ket{g} + C_e\ket{e}$, where $C_g$ and $C_e$ are complex numbers, and quantum state of the whole system for unentangled atoms can be expressed as $\ket{\Psi} = \prod_{k=1}^{N}\ket{\psi}_k$. It then follows that $\ev*{\hat{Q}_M} = N\ev*{\hat{q}_M}$. It can also be seen that $\ev*{\hat{Q}^2_M} = \sum_{k=1}^{N}\hat{q}_{M,k} \sum_{k'=1}^{N}\hat{q}_{M,k'} = N\ev*{\hat{q}^2_M} + N(N-1)\ev*{\hat{q}_M}^2$. Here the first term results from the products of operators corresponding to the same atom, and the second term follows from the product of operators corresponding to a given atom (of which there are $N$) and every other atom (of which there are $N-1$). Let us denote as $\rho \equiv \ev*{\hat{q}_{M}}$ and the corresponding standard deviation of the signal (SDS) as $\Delta\rho \equiv \qty[\ev*{\hat{q}^2_{M}} - \rho^2]^{1/2}$. We also define $\wp \equiv \ev*{\hat{Q}_{M}}$ and the corresponding SDS as $\Delta\wp \equiv \qty[\ev*{\hat{Q}^2_{M}} - \wp^2]^{1/2}$. We thus find the very general result that $\Delta\wp \equiv \qty[N\qty(\ev*{\hat{q}^2_{M}} - \ev*{\hat{q}_{M}}^2)]^{1/2} = \sqrt{N}\Delta\rho$. This, of course, has the rather simple physical meaning that, for unentangled atoms, the total variance (equaling the square of the standard deviation) is the sum of the variances from each atom. Yet, it must be noted that this result only holds when the operator to be measured for the whole system can be viewed as a sum of operators for measuring each atom.

We now address two particular examples of the operator to be measured. First, we consider the case where $\hat{q}_M = \hat{j}_z/\hbar$ ($j = 1/2$), so that $\hat{Q}_M = \hat{J}_z/\hbar$. For each atom, this is equivalent to measuring half the difference in population between the spin-up and spin-down states: $\hat{q}_M = j\qty(\op{e} - \op{g})$. As such, we get $\hat{q}^2_M = j^2\qty(\op{e} + \op{g})$, and for the CRAIN, $\rho = -(1/2)\cos\phi$ and $\wp = -(N/2)\cos\phi$, so that $\Delta\rho = (1/2)\abs{\sin\phi}$ and $\Delta\wp = (\sqrt{N}/2)\abs{\sin\phi}$, yielding QFR$^{-1}_{CRAIN} = \sqrt{N}/\Gamma$. Experimentally, this measurement is the same as that done for the CD-SCAIN, namely measuring the state of each atom, but \textit{the result is very different}, because of the nature of the SC-state.

Next, we consider the case where $\hat{q}_M = j - \hat{j}_z/\hbar$, so that $\hat{Q}_M = J - \hat{J}_z/\hbar$. For each atom, this is equivalent to measuring the population of the spin-down state: $\hat{q}_M = \op{g}$. As such, we get $\hat{q}^2_M = \op{g} = \hat{q}_M$, and for the CRAIN, $\rho = \cos[2](\phi/2)$ and $\wp = N\cos[2](\phi/2)$, so that $\Delta\rho = (1/2)\abs{\sin\phi}$ and $\Delta\wp = (\sqrt{N}/2)\abs{\sin\phi}$, yielding QFR$^{-1}_{CRAIN} = \sqrt{N}/\Gamma$. Experimentally, this CRAIN measurement may appear to be the same as measuring the population of the collective state $\ket{E_0}$, corresponding to the measured operator being $\op{E_0} = \hat{Q}_{M,CSD,0}$, However, \textit{that is not the case}. Indeed, it is easy to see that
\begin{align}
\hat{Q}_{M} & = \sum_{k = 1}^{N}\hat{q}_{M,k} = \sum_{k = 1}^{N}(\op{g})_{k} = J - \hat{J}_z/\hbar \nonumber \\
		   & = J\sum_{m = -J}^{J}\op{E_{J+m}} - \sum_{m = -J}^{J}m\op{E_{J+m}} \nonumber \\
		   & = \sum_{m = -J}^{J-1}(J-m)\hat{Q}_{M,CSD,J+m}
\label{eq:7}
\end{align}
which is a weighted sum of all the operators corresponding to measuring the collective states, excluding the all spin-up state. Eq.~\ref{eq:7} is a \textit{very important expression} that shows the difference between measuring the population of the collective state $\ket{E_0}$ and measuring the population of each atom in the ground state $\ket{g}$.

\subsection{QPD evolutions for odd value of N}

\begin{figure}[h]
\includegraphics[scale=0.28]{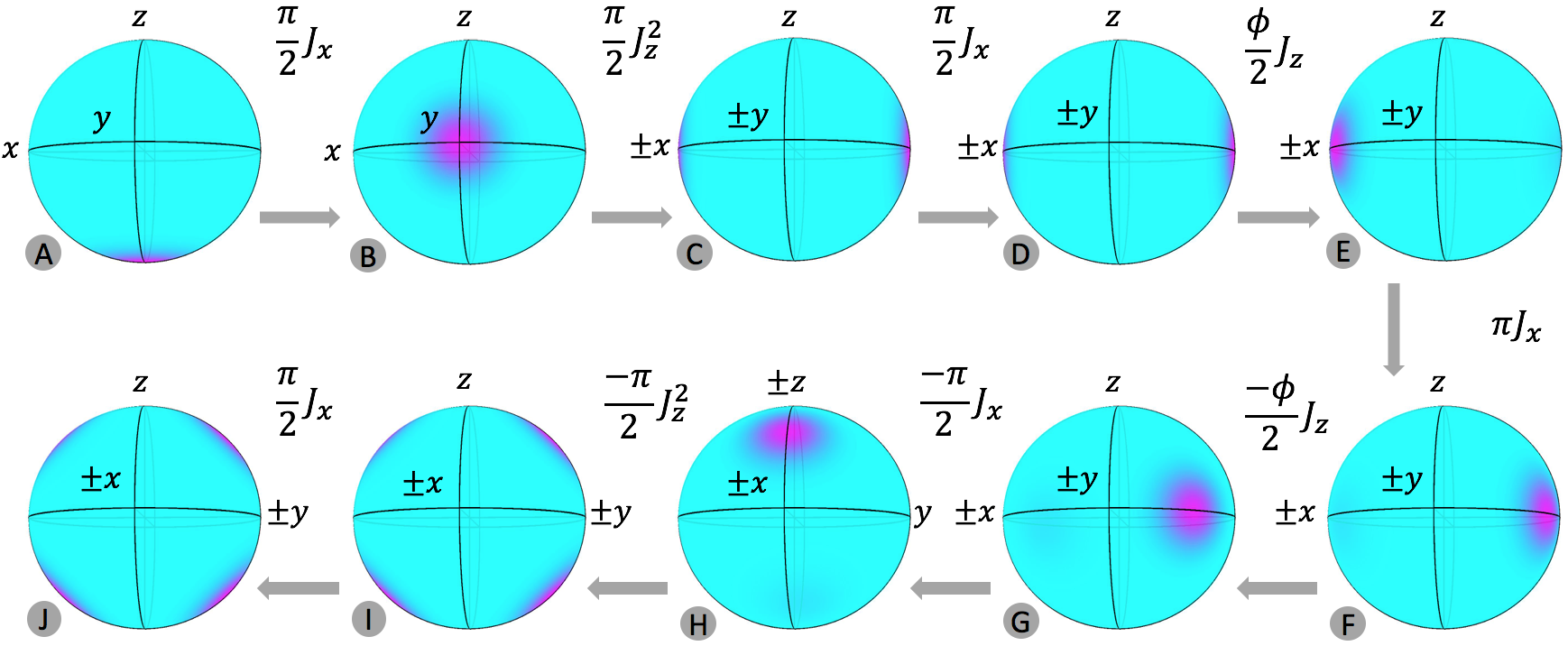}
\caption{The QPDs for different stages of the SCAIN protocol, with $N=41$, $\mu = \pi/2$, $\text{ARA}=\vu{x}$, $\xi = -1$ and $\phi = \pi/4$.}
\label{fig:5}
\end{figure}

Earlier, we showed the QPD evolution for the SCAIN protocol for the case when $N$, the total number of atoms, is even. For comparison, in this subsection we show the QPD evolution for the same protocol for the case when $N$ is odd, as illustrated in ~\ref{fig:5}. All the parameters here are the same as those used to produce the QPD evolution for the even case, except now $N=41$ and $\phi = \pi/4$. 

As mentioned earlier, a very significant difference is observed after the application of the squeezing pulse from time points B to C. Since $N$ is odd, $H_{OAT}$ transforms $\ket{\vu{y}}$ to $(\ket{\vu{x}} + \eta\ket{-\vu{x}})/\sqrt{2}$, where $\eta = i(-1)^{(N+1)/2}$, representing a phase factor with unity amplitude~\cite{Sorensen,Leibfried1,Leibfried2}. It should be noted that the phase factor depends on the super-odd-parity, representing whether $(N+1)/2$ is even or odd; however, the shapes of the fringes, as well as the values of QFR$^{-1}$, for both CSD and CD protocols, are not expected to depend on the value of the super-odd-parity, as we have verified explicitly. This state, illustrated in the QPD at time point C, also represents an SC state, as a superposition of two extremal collective states, but in terms of the x-directed Dicke collective states (XDCSs). If we were to use a protocol where the auxiliary rotation axis (ARA) is the $\vu{y}$ axis, we could produce results similar to what is shown in ~\ref{fig:2}~(b) earlier. However, since we are using the protocol that is designed to produce maximum phase magnification for the case where the ARA is the $\vu{x}$ axis, the result is drastically different. The application of the rotation by $\pi/2$ around the $\vu{x}$ axis from time points C to D leaves the QPD unchanged. The rotation in the first dark zone by an angle of $\phi/2$ around the $\vu{z}$ axis (D to E) moves the QPD in the $x$-$y$ plane on both sides, as shown at time point E. This rotation is inverted by the $\pi$ pulse from E to F. The rotation in the second dark zone by an angle of $-\phi/2$ around the $\vu{z}$ axis (F to G) moves the QPD in the $x$-$y$ plane further on both sides, as shown at time point G. This is followed by a rotation of $-\pi/2$ around the $\vu{x}$ axis from G to H. The unsqueezing pulse turns the QPD distribution into four lobes in the $y$-$z$ plane, as shown at time point I. The final $\pi/2$ pulse rotates this pattern by $90$ degrees, but still with a four-lobed pattern in the $y$-$z$ plane, as shown at time point J. Unlike the case for even values of $N$, it is not easy to write down explicitly the mathematical expression for this final quantum state for an arbitrary value of $\phi$. Instead, we have illustrated the results obtained using numerical simulations.

\subsection{Collective state distributions for both even and odd values of $N$}

For further insight into the behavior of the SCAIN, we also show the population of the collective states corresponding to each stage of the protocol for both even and odd values of $N$. For each case, the set of parameters are the same as those used to generate the QPD plots. 

\begin{figure}[h]
\includegraphics[scale=0.28]{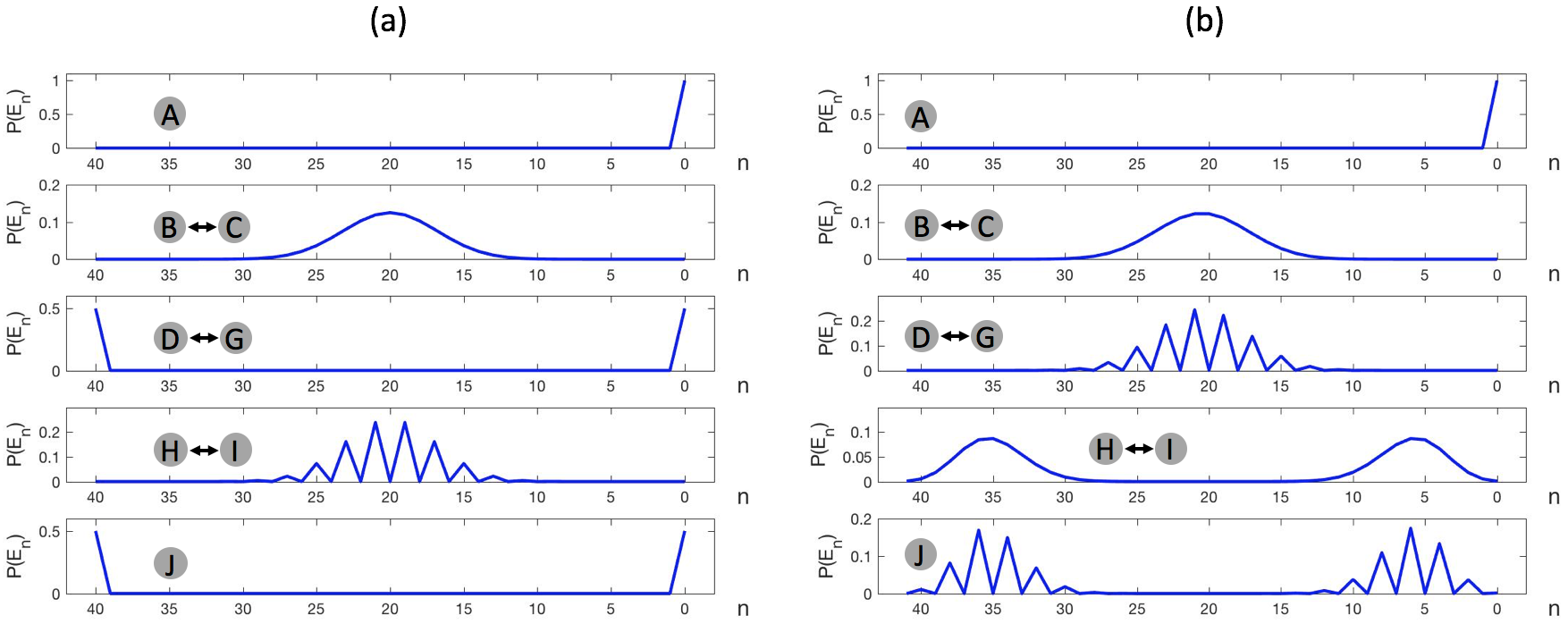}
\caption{Population of the collective states at various stages of the CD-SCAIN protocol: (a) For the case with $N=40$, $\phi=\pi/80$; (b) For the case with $N=41$, $\pi/4$. For both cases, we have $\mu = \pi/2$, $\text{ARA}=\vu{x}$ and $\xi = -1$.}
\label{fig:6}
\end{figure}

Fig.~\ref{fig:6}~(a) corresponds to the case when $N=40$. At the onset, time point A, the system is in the $\ket{E_0}$ state. At time point B, the system is in a coherent spin state , with collective state populations centered around $\sim \ket{E_{N/2}}$ . Perhaps somewhat surprisingly, the distribution of collective states remains unchanged at time point C, after the squeezing pulse, even though in the Bloch sphere it is represented by two lobes on opposite sides. After the auxiliary rotation, at time point D, the system is in a superposition of only two collective states, $\ket{E_0}$ and $\ket{E_N}$, representing the SC state. The distribution of collective states remains unchanged at time points E, F and G. After the corrective auxiliary rotation, at time point H, the distribution returns to a shape with an envelope that is the same as that for a coherent spin state. However, the distribution is modulated, with the depth of modulation determined by the phase shifts accumulated during the two dark zones. This modulated distribution pattern remains unchanged, at time point I, after the unsqueezing pulse. At the final time point J, the system again consists of just two collective states: $\ket{E_0}$ and $\ket{E_N}$. For the particular choice of $\phi$ used here, these populations are equal. However, in general, the ratio of populations for the $\ket{E_0}$ and $\ket{E_N}$ in the final stage depends on the value of $\phi$. When detecting the collective state $\ket{E_0}$, we get a signal that is cosinusoidal, with fringes narrowed by a factor of $N$. As shown in Section ~\ref{Section 2}, we also get fringes with the same factor of narrowing when we detect the atomic states. 

Fig.~\ref{fig:6}~(b) corresponds to the case when $N=41$. The distributions for time points A and B are the same as that for $N=40$. At time point C, the quantum state is different, as can be seen in the QPD plots in Fig.~\ref{fig:5}, with two lobes at the end of the $\pm\vu{x}$ axes on the Bloch sphere. However, the distribution of collective states is still the same as that at time point B. At time point D, after the auxiliary rotation, the QPD remains the same, but the distribution of collective states is now modulated. This distribution remains unchanged at time points E, F and G, despite the phase accumulated in the two dark zones. The modulations disappear at time point H after the application of the corrective auxiliary rotation, and the distribution is split into two distinct lobes. The separation between these two lobes depend on the value of $\phi$. After the unsqueezing pulse, at time point I, the distribution remains the same as that at time point H. The final pulse produces modulations in each lobe. However, it should be noted that, unlike the case of $N=40$, there is no population in either of the extremal collective states. Thus, when detecting the collective state $\ket{E_0}$ , the signal is zero. On the other hand, if the atomic states are detected, the signal as a function of $\phi$ is akin to that of a collective state atomic interferometer (COSAIN) ~\cite{Sarkar3}, although with different amplitudes.

\subsection{Fringe shapes for different values of the squeezing parameter $\mu$}

\begin{figure}[h]
\includegraphics[scale=0.26]{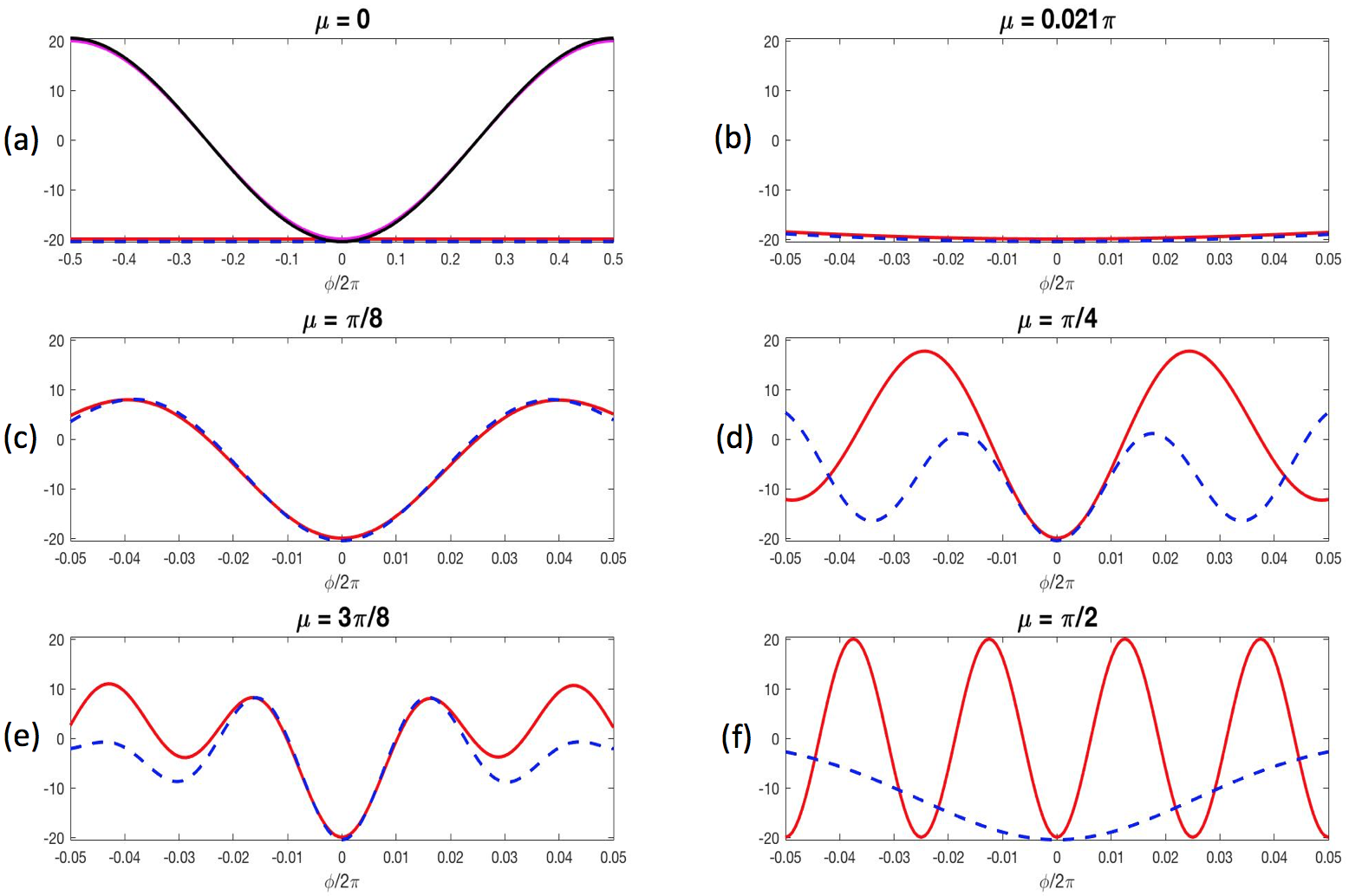}
\caption{Fringe shapes for different values of the squeezing parameter $\mu$, while keeping the rest of the protocol unchanged, for ARA$ =\vu{x}$ and $\xi = -1$. $N=40$ is red while $N=41$ is dashed-blue. (a) $\mu = 0$; (b) $\mu = 0.021\pi$; (c) $\mu = \pi/8$; (d) $\mu = \pi/4$; (e) $\mu = 3\pi/8$; (f) $\mu = \pi/2$. The black line in (a) shows the full fringe of a CRAIN for comparison, and the horizontal spans in (b)-(f) are $10$ times smaller than that in (a).}
\label{fig:7}
\end{figure}

Earlier, we presented the SCAIN protocol primarily for the case of $\mu=\pi/2$, since this is the condition that produces the SC states. However, it is also instructive to consider the behavior of the CD-SCAIN as a function of the squeezing parameter $\mu$, while keeping all other aspects (except $\phi$) of the protocol unchanged. In Fig.~\ref{fig:7}, we illustrate the CD-SCAIN signal, as a function of $\phi$, for different values of $\mu$, for $\text{ARA}=\vu{x}$ and $\xi = -1$. Fig.~\ref{fig:7}~(a) shows the signal for $\mu=0$, where for comparison, we have also shown, as the black line, a full fringe of the CRAIN signal. For increasing values of $\mu$, as shown in Fig.~\ref{fig:7}~(b)-(e), the central fringes become increasingly narrower. It should be noted that for these values of $\mu$, the signals do not have a periodic behavior within the range of $\phi=-\pi$ and $\phi=\pi$. In Fig.~\ref{fig:7}~(f), we show the limiting case of $\mu=\pi/2$. As can be seen, the width of the central fringe remains the same for both odd and even values of $N$ for values of $\mu$ somewhat less than $\pi/2$. In determining the values of $\text{QFR}^{-1}$ for these cases (shown in Fig.~\ref{fig:4} earlier), we have assumed that the interferometer would operate near the central fringe. Thus, the critical differences between the behavior of the odd and even values of $N$ become manifest only when we are very close to or at the value of $\mu=\pi/2$.

\subsection{Determination of average quantum fluctuation of rotation when $N$ can be even or odd with equal probabilities}

In a typical experiment where atoms caught in a magneto-optic trap, for example, and released for atomic interferometry, it is likely that the probability of $N$ to be even or odd would be equal.  We now  determine the average value of $\text{QFR}^{-1}$ for this scenario.  We note first that, for very small rotations, the signal for both even and odd values of N can be approximated as being cosinusoidal, as can be seen from Fig. ~\ref{fig:3}.  Thus, in this limit, we can write the signals as:
\begin{align}
& {{S}_{1}}= -N_1 \text {Cos}^2 ({{{\Omega}_G/2 {\gamma}_1}}) \nonumber \\
& {{S}_{2}}= -N_2 \text {Cos}^2 ({{{\Omega}_G/2 {\gamma}_2}})
\label{eq:7a}
\end{align}
Here, the subscript $1(2)$ refers to the even(odd) case, ${\Omega}_G$ is the rate of rotation, $S$ is the signal, $N$ is the number of atoms, and $\gamma$ is the effective linewidth.  It then follows that $\text{QFR}_1^{-1}=\sqrt{N_1}/\gamma_1 \equiv \alpha_1$ and $\text{QFR}_2^{-1}=\sqrt{N_2}/\gamma_2 \equiv \alpha_2$. Using the subscript $A$ to indicate the average, we can write:
\begin{align}
& S_{A} = (S_1+S_2)/2= -\frac {1}{2} \ \sum_{j = 1}^{2} N_j  \text {Cos}^2 ({{\Omega_G/2 \gamma_j}})  \nonumber \\
&  \frac {\partial S_A}{\partial \Omega_G} \  = \frac {1}{4} \ \sum_{j = 1}^{2} \frac {N_j }{\gamma_j} \  \text {Sin}\qty( \frac{\Omega_G}{\gamma_j}\ ) = \frac {1}{4} \ \sum_{j = 1}^{2} \alpha_j^2 \gamma_j  \text {Sin}\qty( \frac{\Omega_G}{\gamma_j}\ ) \nonumber \\
& \Delta S_{A}^2 =  \frac {1}{8} \ \sum_{j = 1}^{2} N_j  \text {Sin}^2 \qty( \frac{\Omega_G}{\gamma_j}\ )= \frac {1}{8} \ \sum_{j = 1}^{2} \alpha_j^2 \gamma_j^2  \text {Sin}^2 \qty( \frac{\Omega_G}{\gamma_j}\ )
\label{eq:7b}
\end{align}
In the limit of small $(\Omega_G / \gamma_j)$, we thus get, using $  \text {Sin} (\Omega_G / \gamma_j) \approx (\Omega_G / \gamma_j)$ :
\begin{equation}
\begin{split}
\text{QFR}_A^{-1} &=\frac {\abs{\partial S_A /\partial \Omega_G} } {\Delta S_A} \ ={\qty [ \sum_{j = 1}^{2} \frac {\alpha_j^2}{2} \ ]}^{1/2} \\
                              &=\qty[(\text{QFR}^{-1}_1)^2/2 + (\text{QFR}^{-1}_2)^2/2]^{1/2}
\end{split}
\label{eq:7c}
\end{equation}
Thus, the value of the averaged $\text{QFR}^{-1}$ is a factor of $\sqrt {2}$ smaller than the rms of the values of $\text{QFR}^{-1}$ for odd and even values of $N$.  
\subsection{Justification of the dark zone operations}

As mentioned earlier, we have assumed that the phase shift for the SCAIN can be split equally between the two dark zones, and applied operations $e^{-i\frac{\phi}{2}\hat{J}_z}$ ($e^{i\frac{\phi}{2}\hat{J}_z}$) for the first (second) dark zone. These operations can be easily understood in the case of a CRAIN. It can also be easily understood for the case of $\mu=\pi/2$ under the protocol presented here. For an arbitrary value of $\mu$, the quantum state prior to the first (second) dark zone may be distorted in a way so that the concept of two clear trajectories (forming different paths of the Michelson interferometer) may not hold. As such, it may not be obvious whether the application of this operation is valid for such a case. In fact, this operation remains valid under all conditions. Specifically, using an Hamiltonian to represent the Sagnac effect, $H_{SE} = \va{\Omega}_G\vdot(\va{r}\cp\va{p})$, where $\va{r}$ is the position and $\va{p}$ is the momentum of an atom, the phase difference between paths traversed by the $\ket{\uparrow}$ and $\ket{\downarrow}$ components of the $i$-th atom can be accounted for by the operation $e^{-i\Delta\phi\hat{j}_{i,z}}$, where $\Delta\phi = 2m\Omega_G\Delta A/\hbar$, with $\Delta A$ being the differential area enclosed by these paths. Since $\hat{J}_z \equiv \sum_i^N\hat{j}_{i,z}$, it then follows that operations for the evolutions in the dark zones are valid in general.

\section{Schr\"odinger Cat Atomic Clock Using Conventional Detection}
\label{Section 4}

As described in Ref.~\cite{Sarkar3}, the combination of one-axis-twist squeezing (OATS), rotation, unrotation, unsqueezing and collective state detection can also be used to realize a Schr\"odinger Cat Atomic Clock (SCAC) with HL sensitivity. We will refer to this as the CSD-SCAC. Earlier, we mentioned that such a SCAC with HL sensitivity can also be realized when conventional detection of atomic states is employed. We will refer to this as the CD-SCAC. In this section we present the results obtained for the CD-SCAC, and comparison thereof with the CSD-SCAC.

\subsection{Conventional atomic clock and Collective State atomic clock}
In order to describe how the CSD-SCAC and the CD-SCAC work, it's useful to review briefly some details about the conventional atomic clock (CAC)  as well as the collective state atomic clock (COSAC)~\cite{Kim}. Here we consider a system where the ground states, $\ket{1}$ and $\ket{2}$ of a three-level atom interact with an excited state $\ket{3}$ via two copropagating laser beams. One of the beams is detuned from resonance by $\delta_1$ and has a Rabi frequency $\Omega_1$; this couples $\ket{1}$ to $\ket{3}$. The second beam is detuned from resonance by $\delta_2$ and has a Rabi frequency $\Omega_2$; this couples $\ket{2}$ to $\ket{3}$. For $\delta \gg \Omega_1$, $\Omega_2$, $\Gamma$, where $\delta \equiv (\delta_1 + \delta_2)/2$ and $\Gamma$ is the excited state decay rate, the system can be modeled as an effective two level system, consisting of states $\ket{1}$ and $\ket{2}$, excited by a traveling wave with a Rabi frequency $\Omega = \Omega_1\Omega_2/(2\delta)$, and detuning $\Delta \equiv \delta_1 - \delta_2$. For simplicity, we assume $\Omega_1 = \Omega_2$, and $\Delta \ll \delta$, so that $\delta_1 \simeq \delta_2$. Under this condition, the light-shifts experienced by states $\ket{1}$ and $\ket{2}$ are essentially the same, and do not affect the equation of motion~\cite{Shahriar3}. For more general cases, it is possible to incorporate any differences in the light shifts into the definition of $\Delta$. Just as in the case of the SCAIN discussed earlier, we denote states $\ket{1}$ and $\ket{2}$ as being the pseudo-spin states $\ket{\downarrow}$ and $\ket{\uparrow}$, respectively. It should be noted that this is formally equivalent to a conventional microwave atomic clock that couples state $\ket{1}$ to state $\ket{2}$. However, since a Raman transition is needed for the CSD protocol, we choose to describe it here as a Raman clock. In practice, for both the CSD and the CD protocols, all results presented here would remain valid for a conventional microwave excitation, which is preferable because a Raman clock may suffer from fluctuations in light shifts.

In a conventional Raman Ramsey atomic clock, which is equivalent to a CAC, an ensemble of $N$ effective two-level atoms is first prepared in a coherent spin state, denoted as $\ket{-\vu{z}} \equiv \ket{E_0} = \prod_{k = 1}^{N}\ket{\downarrow_k}$. The first $\pi/2$ pulse produces a rotation about the $\vu{x}$ axis. During the interval, $T_D$, before the second $\pi/2$ pulse, each atom acquires a phase $\phi=2\pi fT_D$, where $f = \Delta/2\pi$ is the (two-photon) detuning of the clock (in Hertz). Application of the second $\pi/2$ pulse around the $\vu{x}$ axis produces the final state, which, for each atom, can be expressed, ignoring an overall phase-factor, as:
\begin{equation}
\begin{split}
\ket{\Psi} &=e^{-i\frac{\pi}{2}\hat{J}_x} e^{-i\phi\hat{J}_z} e^{-i\frac{\pi}{2}\hat{J}_x} \ket{-\vu{z}} \\
	       &=\prod_{k = 1}^{N}\frac{1}{2}\{(1 - e^{i\phi})\ket{\downarrow_k} - i(1 + e^{i\phi})\ket{\uparrow_k}\}
\end{split}
\label{eq:8}
\end{equation}

In a CAC, typically the signal is a measure of the population of $\ket{\uparrow}$, given by $S_{CAC} = J + \ev*{\hat{J}_z} = N\cos[2](\phi/2)$. The associated quantum projection noise is $\Delta S_{CAC} = \Delta \hat{J}_z = \sqrt{N/4}\abs{\sin\phi}$. The stability of the clock is attributed to the quantum fluctuation in frequency (QFF), analogous to the QFR described earlier for a rotation sensor based on an atomic interferometer. This can be expressed as $\text{QFF} = \Delta f|_{CAC} = \Delta(\hat{J}_z)/\partial_f\ev*{\hat{J}_z} = (2\pi T_D\sqrt{N})^{-1}$, where $\partial_f \equiv \pdv*{}{f}$. This can also be written as $\Delta f|_{CAC} = \gamma/\sqrt{N}$, where $\gamma = 1/(2\pi T_D)$ is the effective linewidth. This is, of course, the SQL value of the QFF.

In a COSAC, however, the signal is a measure of the population of one of the extremal collective states and is given by $S_{COSAC} = \ev*{\hat{Q}} = \cos[2N](\phi/2)$, where $\hat{Q} \equiv \op{E_N}$. This signal shows a $\sqrt{N}$-fold reduction in fringes compared to that of a CAC, which can be explained as follows. The first $\pi/2$ pulse couples the initial state $\ket{E_0}$ to $\ket{E_1}$, which in turn is coupled to $\ket{E_2}$ and so on, effectively causing the ensemble to split into $N+1$ states. During the dark zone, the $n$-$th$ collective state $\ket{E_n}$ picks up a phase $e^{-in\phi}$. When the ensemble interacts with the last $\pi/2$ pulse, each of the collective states interferes with the rest of the collective states. The COSAC can thus be viewed as the aggregation of interference patterns due $\binom{N+1}{2}$ CAC's working simultaneously~\cite{Kim}. The narrowest constituent signal fringes are derived from interferences between states with the largest difference in phase, i.e. $\ket{E_0}$ and $\ket{E_N}$; the width of this fringe is $\gamma/N$. The width of the rest of the signal components range from $\gamma$ to $\gamma/(N-1)$. The signal, which is the measure of population of $\ket{E_N}$, is the result of the weighted sum of all the pairwise interferences, with a width of $\gamma/\sqrt{N}$. However, the system acts as a single particle, which reduces the effective SNR by the factor of $\sqrt{N}$. As a result, we have shown that the QFF for the COSAC is essentially the same as that for the CAC~\cite{Kim}. 

From the analyses above, it follows that if the evolution of the system could be restricted to just the two extremal Dicke states (namely, $\ket{E_0}$ and $\ket{E_N}$) during the dark zone evolution, the fringes would be narrowed by a factor of $N$ compared to those of the CAC. In that case, the QFF would be enhanced by a factor of $\sqrt{N}$, thus reaching the HL sensitivity. As noted earlier, the process of OATS indeed can be used to create just such a Schr\"odinger Cat (SC) state if the degree of squeezing is chosen to be $\mu=\pi/2$, and an auxiliary rotation of $\pi/2$ is applied along a particular axis after the squeezing pulse. The resulting clock is then referred to as the SCAC.

\subsection{The complete protocol for the SCAC}

\begin{figure}[h]
\includegraphics[scale=0.34]{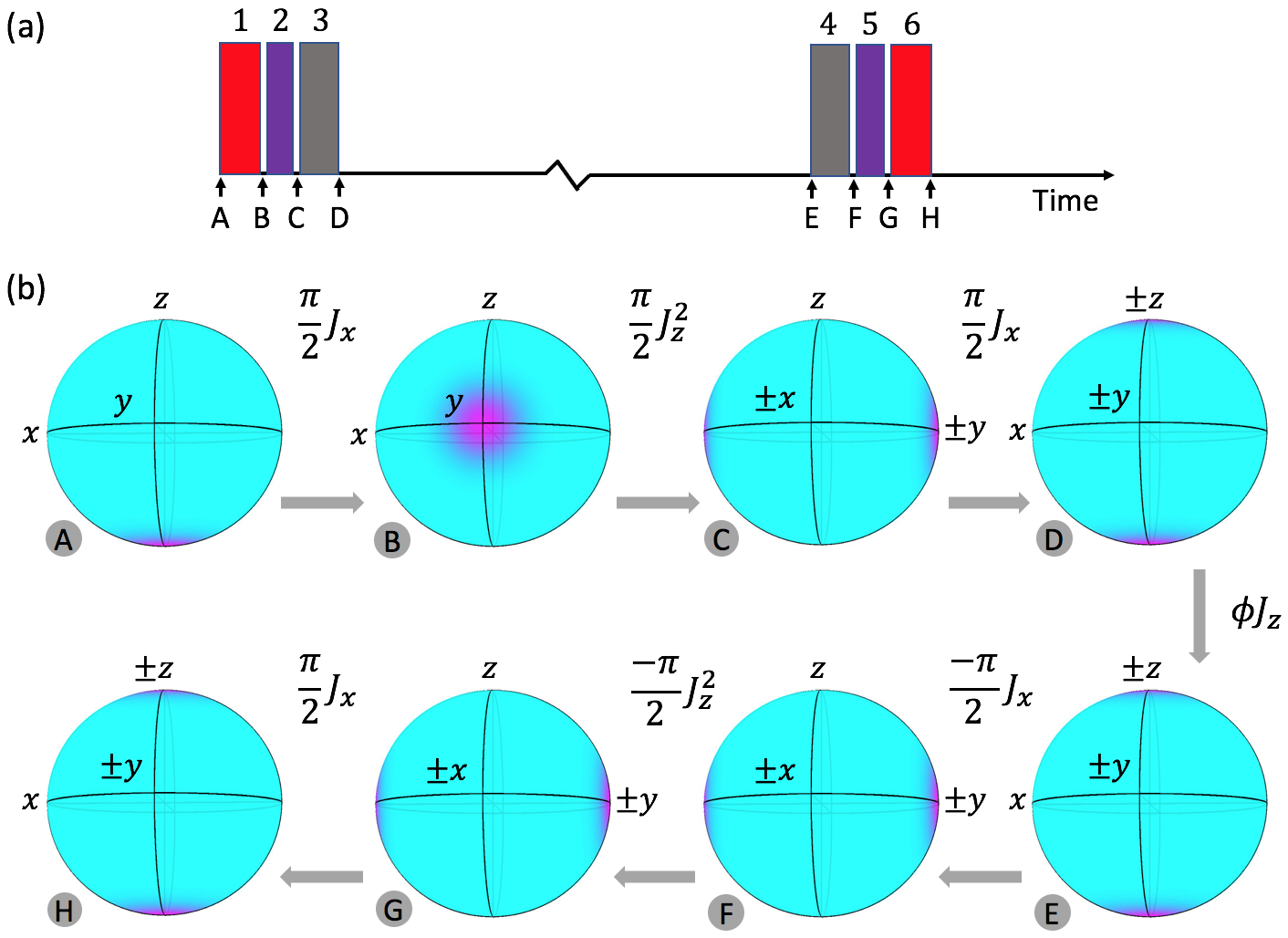}
\caption{(a) Schematic illustration of the protocol employed for Schroedinger Cat Atomic Clock (SCAC). (b) The QPDs at different stages of the protocol, for $N=40$, $\mu=\pi/2$, $\text{ARA}=\vu{x}$, $\xi=-1$ and $\phi=0.5\pi/N$.}
\label{fig:8}
\end{figure}

Just as in the case of the SCAIN, the exact effects of the protocol depend on a set of parameters such as the value (and parity) of $N$, the squeezing parameter $\mu$ for the OATS, the auxiliary rotation axis (ARA, which can be $\vu{x}$ or $\vu{y}$ ) around which to implement the rotation, the corrective rotation sign $\xi$ which can take values of $\pm1$ corresponding to redoing or undoing the first auxiliary rotation, and lastly the dark zone phase shift $\phi$. The protocol illustrated in Fig.~\ref{fig:8}~(a) corresponds to the ARA chosen to be the $\vu{x}$ axis. The process starts by applying a $\pi/2$ pulse around the $\vu{x}$ axis. This is followed by the application of OATS, corresponding to a rotation around the $\vu{z}$ axis by an angle of $\mu J_z$, with $\mu=\pi/2$. The next step is an auxiliary rotation of $\pi/2$ around the $\vu{x}$ axis. The ensuing evolution in the dark zone corresponds to a rotation by $\phi$ around the $\vu{z}$ axis, where $\phi=2\pi fT_D$. This is now followed by another auxiliary rotation around the $\vu{x}$ axis, by an angle of $\xi\pi/2$. This is followed by an unsqueezing pulse, which corresponds to a rotation around the $\vu{z}$ axis by an angle of $-\mu J_z$, with $\mu=\pi/2$. Finally, the protocol ends with the application of the final $\pi/2$ pulse around the $\vu{x}$ axis. Mathematically, for this choice of the ARA, the whole protocol can thus be expressed as:
\begin{multline}
\ket{\Psi}_f = e^{-i\frac{\pi}{2}\hat{J}_x} e^{i\mu\hat{J}_z^2} e^{-i\xi\frac{\pi}{2}\hat{J}_x} e^{-i\phi\hat{J}_z} e^{-i\frac{\pi}{2}\hat{J}_x} \\
e^{-i\mu\hat{J}_z^2} e^{-i\frac{\pi}{2}\hat{J}_x} \ket{-\vu{z}}
\label{eq:9}
\end{multline}

In Fig.~\ref{fig:8}~(b), we show the evolution of the quantum states on a Bloch sphere, using the QPD, for an even value of $N=40$, with $\mu=\pi/2$, $\xi=-1$ and $\phi=0.5\pi/N=\pi/80$. In illustrating the nature of the QPD at various stages of the protocol, we have used different orientations, as needed. At the onset of the process (time point A), the system is assumed to be in the state $\ket{E_0} = \ket{-\vu{z}}$, which is a coherent spin state. After the first $\pi/2$ rotation around the $\vu{x}$ axis (time point B), it is in state $\ket{\vu{y}}$. After the squeezing pulse, the state (time point C) is split between two coherent spin states, and can be expressed as $(\ket{\vu{y}} - \eta\ket{-\vu{y}})/\sqrt{2}$, , where $\eta = i(-1)^{N/2}$, representing a phase factor with unity amplitude. This factor depends on the super-even-parity , representing whether $N/2$ is even or odd. However, the shapes of the fringes, as well as the values of $\text{QFF}^{-1}$, for both CSD and CD protocols, are not expected to depend on the value of the super-even-parity, as we have verified explicitly. Application of the auxiliary rotation of $\pi/2$ around the $\vu{x}$ axis transforms this state to $(\ket{-\vu{z}} + \eta\ket{\vu{z}})/\sqrt{2}$. This (time point D) represents the desired SC state, as a superposition of the two extremal states of the ZDCS manifold: $(\ket{E_0} + \eta\ket{E_N})/\sqrt{2}$.

During the dark zone, the phase shift causes a rotation by an angle of $\phi$ around the $\vu{z}$ axis, for each atom. The state after the dark zone can be expressed as $e^{-i\phi\hat{J}_z} \qty(\ket{E_0} + \eta\ket{E_N})/\sqrt{2}$. Since both $\ket{E_0}$ and $\ket{E_N}$ are eigenstates of the $\hat{J}_z$ operator, with eigenvalues (assuming  $\hbar = 1$) of $-N/2$ and $N/2$ respectively, this state can be expressed as $\qty(e^{i\phi N/2} \ket{E_0} + e^{-i\phi N/2} \eta\ket{E_N})/\sqrt{2}$. The resulting QPD, shown at time point E of Fig.~\ref{fig:8}~(b), remains unchanged, but the quantum state incorporates these phase accumulations. In order to reveal the interference magnified by the factor of $N$, it is necessary to apply first another auxiliary rotation, by an angle of $\xi\pi/2$ around the $\vu{x}$ axis. The QPD resulting from the case for $\xi=-1$ is shown at time point F. It is then necessary to apply the unsqueezing pulse, by an angle of $-\mu\hat{J}_z$, with $\mu=\pi/2$. The QPD of the resulting state is shown at time point G. Finally, it is necessary to apply one more rotation around the $\vu{x}$ axis, by an angle of $\pi/2$. The QPD for the final state is shown at time point H.

It is easy to show that, for this case, the final state can be expressed as $\ket{\Psi}_f = \eta\cos(N\phi/2)\ket{E_N} + \sin(N\phi/2)\ket{E_0}$. For the particular value of $\phi$ (which is $0.5\pi/N$) used in generating the QPDs, the final state is $\qty(\eta\ket{E_N} + \ket{E_0})/\sqrt{2}$. If the population of $\ket{E_N}$ were detected, the signal would be expressed as $\cos[2](N\phi/2)$, with fringes that are a factor of $N$ narrower than that for the CAC. This is the CSD-SCAC discussed in Ref.~\cite{Sarkar1}. Here we show that the same results hold even if the CD process is used, thus realizing the CD-SCAC.

\subsection{Signal fringes for the CD-SCAC}

\begin{figure}[h]
\includegraphics[scale=0.305]{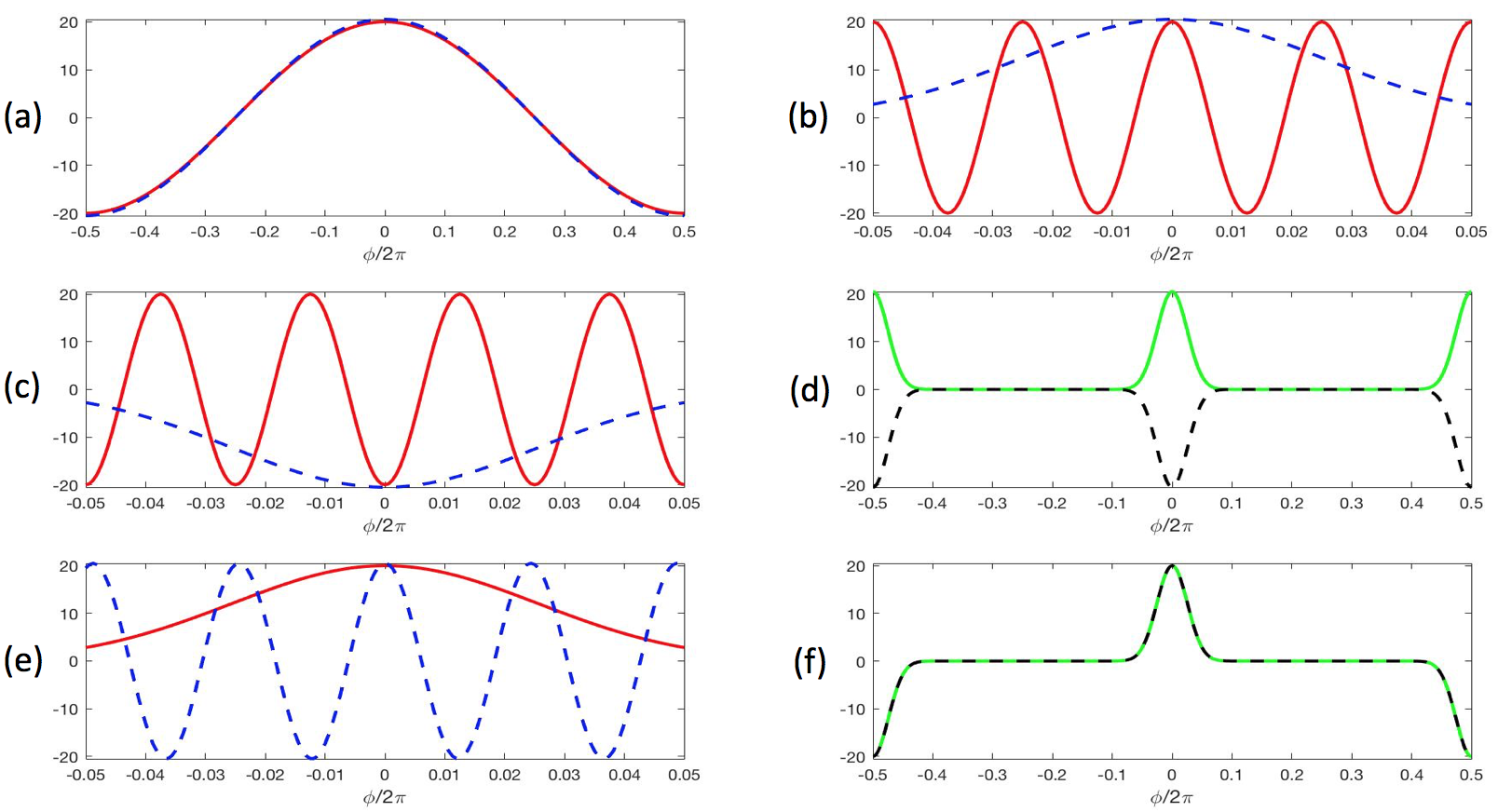}
\caption{Signals corresponding to detection of $\ev*{\hat{J}_z/\hbar}$, as a function of $\phi$. $N=40$ is red while $N=41$ is dashed-blue. (a) Fringes for a CAC for comparison; (b) CD-SCAC with ARA$ =\vu{x}$ and $\xi = -1$; (c) CD-SCAC with ARA$ =\vu{x}$ and $\xi = +1$; (d) Zoomed-out plots for $N=41$ with $\xi = -1$ in green and $\xi = +1$ in black, for CD-SCAC with ARA$ =\vu{x}$; (e) CD-SCAC with ARA$ =\vu{y}$ and $\xi = \pm1$; (f) Zoomed-out plots for $N=40$ with $\xi = -1$ in green and $\xi = +1$ in black, for CD-SCAC with ARA$ =\vu{y}$. Here, $\mu=\pi/2$ for all cases, except in (a) which has no squeezing. Also note the horizontal spans in (b), (c) and (e) are $10$ times smaller than those in (a), (d) and (f).}
\label{fig:9}
\end{figure}

In Fig.~\ref{fig:9}, the signal fringes for the CD-SCAC are plotted as a function of $\phi$ (red for $N=40$ and dashed-blue for $N=41$). For reference, we show in Fig.~\ref{fig:9}~(a) the signal corresponding to one full fringe of a CAC. For the remainder of the figures, $\mu=\pi/2$.

Fig.~\ref{fig:9}~(b) shows the signal for ARA$=\vu{x}$ and $\xi=-1$. Here, the horizontal span of $\phi$ is smaller by a factor of $10$. Consider first the signal for even $N$, in red, which shows $4$ full fringes. This corresponds to a phase magnification by a factor of $N=40$. Since the signal magnitude is the same as that for a CAC, one might be tempted to think that because of this phase magnification, the value of the QFF$^{-1}$ for the CD-SCAC should be higher than that of a CAC by a factor of $N$. However, the standard deviation of the signal (SDS) for the CD-SCAC signal is larger than that for a CAC by a factor of $\sqrt{N}$ (This can be shown analytically in the same way as we had shown, in Section~\ref{Section 3}, how the SDS for the CD-SCAIN is larger than that for a CRAIN by a factor of $\sqrt{N}$) . As such, the net enhancement in the value of the QFF$^{-1}$ is by a factor of $\sqrt{N}$, corresponding to HL sensitivity. 

Consider next the signal for odd $N$, in dashed-blue, which shows a much smaller variation as a function of $\phi$. This same signal is shown again by the green line in Fig.~\ref{fig:9}~(d), but for a much larger range of $\phi$, matching that of a full fringe for a CAC. Thus, the signal for odd values of $N$ is similar to that for a Fabry-Perot resonator, with the width of the central fringe narrowed by a factor of $\sim$$\sqrt{N}$. As such, this signal is analogous to what is found for the COSAC, as detailed in Ref.~\cite{Kim}, with the exception that, in the case of the CD-SCAC, the fringe amplitude is $N/2$, while for the COSAC it is $1$. Again due to the increased SDS, the sensitivity of the CD-SCAC for this case is the same as that for a CAC and the COSAC.

Fig.~\ref{fig:9}~(c) shows the CD-SCAC signal for ARA$ =\vu{x}$ and $\xi = +1$. As expected, in this case the fringes for both even (red) and odd (dashed-blue) values of $N$ are flipped around the zero value. The signal for the odd value of $N$ is shown again by the dashed black line in Fig.~\ref{fig:5}~(d) on a scale where the span of $\phi$ is the same as that for a full fringe of the CAC, again showing the Fabry-Perot type resonance, reduced in width by a factor of $\sim$$\sqrt{N}$. The values of QFF$^{-1}$, and therefore the sensitivities, are the same as those for the case shown in Fig.~\ref{fig:9}~(b).

In Fig.~\ref{fig:9}~(e), we show the signal for a variant of the protocol where ARA$ =\vu{y}$ and $\xi = \pm1$. For this protocol, the behaviors for odd (dashed-blue) and even (red) values of $N$ are essentially reversed. However, for this value of the ARA, we find that the signals are the same for both values of $\xi$. In Fig.~\ref{fig:9}~(f), we show the signal, for the odd value of $N$, on a scale where the span of $\phi$ is the same as that for a full fringe of the CAC, again showing the Fabry-Perot type resonance, reduced in width by a factor of $\sim$$\sqrt{N}$.

\subsection{QFF$^{-1}$ for the CD-SCAC}

\begin{figure}[h]
\includegraphics[scale=0.36]{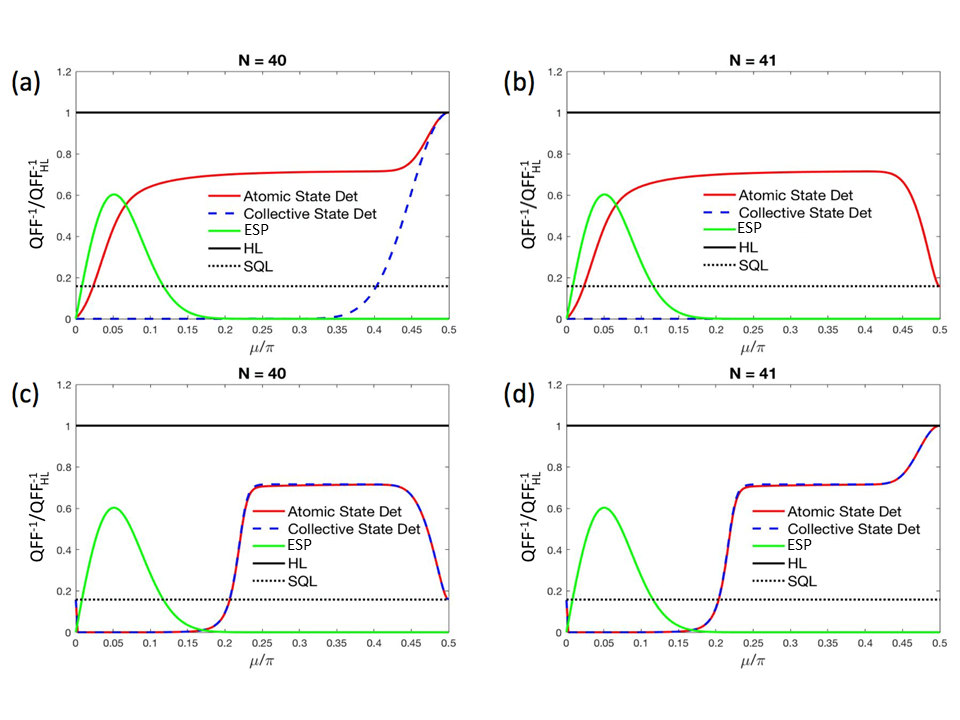}
\caption{Illustration of QFF$^{-1}$ for different cases, as a function of the squeezing parameter $\mu$, normalized to the HL (solid black line). (a) The case for even $N=40$, with ARA as $\vu{x}$; (b) The case for odd $N=41$, with ARA as $\vu{x}$; (c) The case for even $N=40$, with ARA as $\vu{y}$; (d) The case for odd $N=41$, with ARA as $\vu{y}$. The dotted black line shows the SQL. Red is for CD-SCAC, dashed-blue for CSD-SCAC and green for the ESP case. For all cases shown, $\xi=+1$.}
\label{fig:10}
\end{figure}

In Fig.~\ref{fig:10}, we illustrate the behavior of QFF$^{-1}$, as a function of $\mu$, with $\xi=+1$, for different choices of parameters for the CD-SCAC, along with a comparison with the CSD-SCAC and the Echo Squeezing Protocol (ESP)~\cite{Davis,Hosten2}. In each case, the QFF$^{-1}$ is normalized to the QFF$^{-1}_{HL}$ for $N=40$, indicated as the solid black line. The dashed black line shows the QFF$^{-1}_{SQL}$ for $N=40$.

Fig.~\ref{fig:10}~(a) corresponds to $N=40$, with ARA being the $\vu{x}$ axis. Here, the red line corresponds to the CD-SCAC, and the dashed-blue line is for the CSD-SCAC. For $\mu=\pi/2$, we see that the sensitivity for both CD and CSD protocols yield the HL sensitivity. This sensitivity is reached due to an amplification of phase by a factor of $N$, and a concomitant increase in the standard deviation of the signal (SDS) by a factor of $\sqrt{N}$.  Fig.~\ref{fig:10}~(b) is the same as Fig.~\ref{fig:10}(a), except that $N=41$. In this case, for the CD-SCAC with $\mu=\pi/2$, there is a phase amplification, manifested as a Fabry-Perot like fringe around $\phi=0$ which is narrowed by a factor of $\sim$$\sqrt{N}$, along with an increase in the standard deviation of the signal (SDS) by a factor of $\sim$$\sqrt{N}$. The difference between the even and odd cases disappears when the value of $\mu$ is reduced below a threshold value of $\sim0.45\pi$. There is a range of values of the squeezing parameter ($0.2\pi\le\mu\le0.45\pi$) over which the normalized value of QFF$^{-1}$ is $\sim0.71$ for the CD-SCAC. We have verified that this plateau ratio between QFF$^{-1}$ and QFF$^{-1}_{HL}$ remains unchanged when $N$ is increased or decreased. We also see that, for this choice of the ARA, the behavior of the CSD-SCAC is drastically different. Specifically, for odd values of $N$, the QFF$^{-1}$ is strictly zero for all values of the squeezing parameter, and for even value of $N$, the QFF$^{-1}$ drops to zero quickly for $\mu<\pi/2$.

Fig.~\ref{fig:10}~(c) and Fig.~\ref{fig:10}~(d) are similar to Fig.~\ref{fig:10}~(a) and Fig.~\ref{fig:10}~(b), respectively, but with the ARA chosen to be the $\vu{y}$ axis. In this case, it should be noted that the behavior of the CD-SCAC and the CSD-SCAC are essentially the same, except for a small range of value of $\mu$ around $0.05\pi$. We also note that, for this choice of the ARA, the HL sensitivity is reached for odd values of $N$. Finally, in each of these four cases, we have used the green line to show the corresponding sensitivity achievable under the ESP.

So far, we have presented the value of QFF$^{-1}$ separately for odd and even values of $N$. In certain cases, such as for a magnetometer using nitrogen-vacancy color centers in diamond, where it is possible to operate with a fixed parity of $N$, the values of QFF$^{-1}$ for a given parity is relevant. For other situation, such as a clock using atoms cooled in a magneto-optical trap and released for interrogation, it is necessary to consider the effect of averaging over the two parities. Generalizing the result shown in Eqn.~\ref{eq:7c} for the case of the clock, we see that the average value in this case is given by $\text{QFF}^{-1}_{AVE} = \qty[(\text{QFF}^{-1}_{EVEN})^2/2 + (\text{QFF}^{-1}_{ODD})^2/2]^{1/2}$. Using this result, we can reach the following conclusions, assuming $N\gg1$. If $\text{QFF}^{-1}_{EVEN} = \text{QFF}^{-1}_{HL}$ and $\text{QFF}^{-1}_{ODD} = 0$, then $\text{QFF}^{-1}_{AVG} = \text{QFF}^{-1}_{QHL}$, where we define QFF$^{-1}_{QHL} \equiv \text{QFF}^{-1}_{HL}/\sqrt{2}$ . Similarly, if $\text{QFF}^{-1}_{EVEN} = \text{QFF}^{-1}_{HL}$ and $\text{QFF}^{-1}_{ODD} = \text{QFF}^{-1}_{SQL}$, then $\text{QFF}^{-1}_{AVG} \cong \text{QFF}^{-1}_{QHL}$. Finally, if $\text{QFF}^{-1}_{EVEN} = \text{QFF}^{-1}_{QHL}$ and $\text{QFF}^{-1}_{ODD} = \text{QFF}^{-1}_{QHL}$, then $\text{QFF}^{-1}_{AVG} = \text{QFF}^{-1}_{QHL}$.

\section{Experimental Considerations}
\label{Section 5}

In Ref. ~\cite{Sarkar1}, we described in detail a set of specific steps for realizing the CSD-SCAIN, employing atoms released from a magneto-optic trap.  Here, we first describe how the experiment would be simplified considerably due to the use of conventional detection instead of collective state detection.  Then, we analyze in detail the implementation of one axis twist spin squeezing using an optical cavity, taking into account the effect of cavity decay and spontaneous emission. All discussions here are in the context of the CD-SCAIN; however, the findings apply equally well to the CD-SCAC.

\subsection{Experimental Simplification for CD-SCAIN Compared to CSD-SCAIN}

In this subsection, we review briefly the proposed scheme for implementing the CSD technique, and show how the SCAIN protocol can be greatly simplified experimentally by switching from CSD to CD. The complete experimental proposal for realizing the CSD technique is detailed in section IV of Ref.~\cite{Sarkar3}, where a null-detection scheme is employed to measure populations of one of the extremal Dicke collective states. The probe is one of the two counter-propagating Raman beams, which will induce Raman transitions within the atomic ensemble unless it is in the desired extremal collective state. As a result, there will be photons emitted corresponding to the other leg of the Raman transition. The probe and the emitted photons will be combined and sent to a high speed detector, which produces a dc voltage along with a beat signal with a beat frequency the same as that of the frequency synthesizer used to generate the two Raman beams but with an unknown phase. To extract the amplitude, the beat signal is bifurcated and one part is multiplied by the frequency synthesizer signal, while the other is multiplied by the frequency synthesizer signal phase shifted by $90$ degrees. The signals are then squared before being recombined and sent through a low-pass filter to derive the dc voltage. This dc voltage is proportional to the number of scattered photons. A lower limit is set for the voltage reading and any values recorded above it will indicate the presence of emitted photons. If no photon is emitted, the voltage will read below the limit, indicating that the ensemble is in the desired extremal collective state; otherwise at least one photon will be emitted and the ensemble is in other collective states. This process is then repeated many times for a given value of $\phi$. The fraction of events where no photons are detected will correspond to the signal for this value of $\phi$. This process is then repeated for several values of $\phi$, producing the signal fringe for a CSD-SCAIN.

In contrast, the CD technique can be easily realized by coupling one of the two ground states involved in the Raman transition to some upper states of the atom and collecting the fluorescence with a photodetector with high quantum efficiency, thus avoiding the need for the aforementioned heterodyning and quadrature measurements. Moreover, the CSD technique requires an additional ring cavity to increase the optical density in order to enhance the signal (see section V of Ref.~\cite{Sarkar3} for more details), which is not the case for the CD technique. All these factors taken into account, the CD version of the SCAIN protocol will be significantly simpler to implement experimentally.  It should be noted that even though the CSD protocol is experimentally more challenging and more sensitive to excess noise, it may be better suited for some applications, such as the realization of a matter wave clock with ultra-high Compton frequency~\cite{Lan}.  For applications such as testing of the Penrose-Diosi theory of gravitationally induced decoherence~\cite{Diosi1,Diosi2,Penrose1,Marshall,Penrose2}, either protocol can be used.
 
\subsection{One Axis Twist Squeezing: Effect of Non-idealities}
  
\begin{figure}[h]
\includegraphics[scale=0.30]{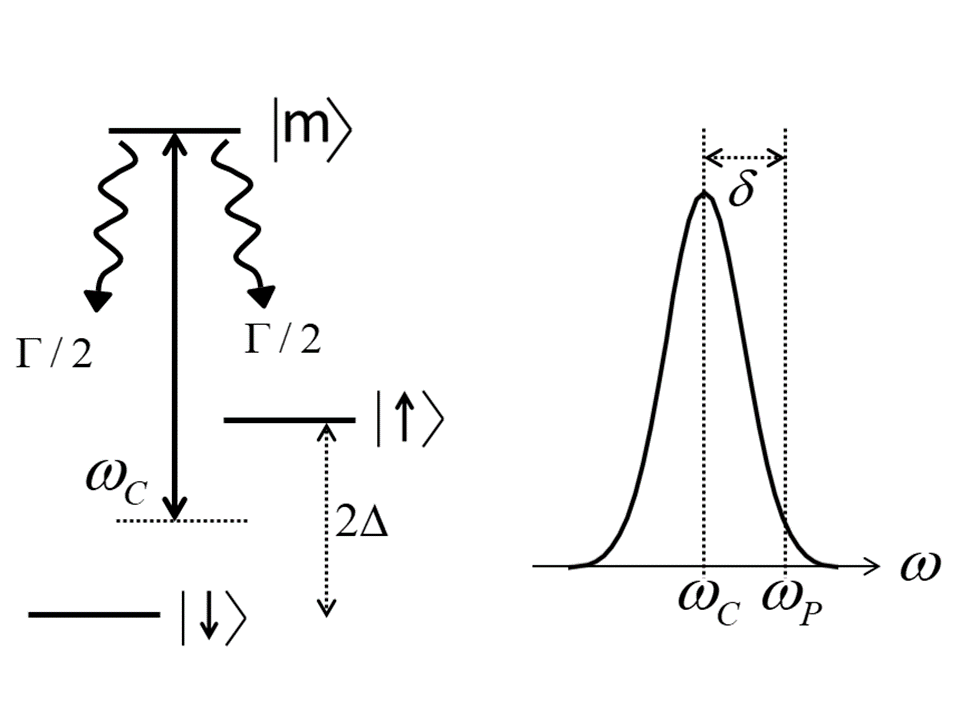}
\caption{Illustration of the scheme considered for one axis twist squeezing, using a three level system, where the two low-lying states are metastable, and represent the spin-up and spin-down states.  These states are coupled optically to an intermediate state, which decays to the ground states via spontaneous emission, at the same rate.  The cavity is tuned to the center of the two transitions, with an equal an opposite amount of detuning with respect to each transition.  The probe laser transmitted to the cavity is detuned away from resonance.  Reversing the sign of the probe detuning causes anti-squeezing.}
\label{fig:11}
\end{figure}

In this subsection, we consider a specific implementation of the one axis twist  squeezing (OATS)  process, tuned to the critical value needed for generating the Schr\"odinger Cat (SC) state needed for the CD-SCAIN, and investigate the effects of various non-idealities, such cavity decay and spontaneous emission.  There are several experimental schemes for realizing one-axis-twist squeezing~\cite{Takeuchi,Schleier,Leroux,Leroux2,Hosten1,Davis,Hosten2,Wang,Liu,Gil}. For concreteness, we consider here the approach based on cavity feedback dynamics~\cite{Schleier,Leroux,Leroux2,Hosten1,Davis,Hosten2}. In this approach, a probe is passed through a cavity, at a frequency that is tuned halfway between the two legs of a $\Lambda$ transition in which the spin-up and spin-down states are coupled to an intermediate state. The cavity is tuned to be below resonance for the probe. The energy levels of the spin-up and spin-down states are light shifted due to the probe, in opposite directions. The resulting dispersion shifts the cavity resonance frequency by an amount that is proportional to $J_z$, the z-component of the total spin for all atoms. The intra-cavity probe intensity changes linearly with this cavity shift, since it is on the side of the resonance, thus affecting the light-shifts. The net result is an energy shift for all the atoms that is proportional to the square of $J_z$, so that the interaction Hamiltonian can be expressed as $H_{OATS} =\hbar \chi J_z^2$, where $\chi$ is a parameter that determines the strength of the squeezing process. Changing the sign of the probe detuning with respect to the cavity resonance reverses the sign of the Hamiltonian, thus producing unsqueezing.  For an OATS interaction time of $\tau$, the characteristic strength for the OATS process is given by $\mu \equiv \chi \tau$, as noted earlier.  The basic scheme for OATS using cavity feedback dynamics is illustrated schematically in Fig.~\ref{fig:11}.  Here, we have denoted the excited state as $\ket {m}$, and the energy separation between the spin-up and spin-down states as $2 \hbar \Delta$, so that the magnitude of the detuning for the cavity mode with respect to either ground state is $\Delta$.

In what follows, we first derive the effective Hamiltonian for the OATS process, while also taking into account the effects of dissipative processes.  This analysis follows steps similar to those found in the supplement of ~\cite{Davis}.  However, we repeat briefly the essential steps since our notations are different, and  spell out some of the steps not explicitly shown there; furthermore, there are some small, although non-critical, discrepancies between the results reported there and what we find.   For specificity, we consider  $^{87}$Rb as the atomic medium, with the spin-up state corresponding to the $5S_{1/2},F=2,m_F=0$ Zeeman sublevel, and the  spin-down state corresponding to the $5S_{1/2},F=1,m_F=0$ Zeeman sublevel.  The intermediate state is assumed to be the $5P_{3/2}$ manifold.  We also assume the matrix element for the coupling to the intermediate state to be the same for both spin-up and spin-down states; in practice, a detailed numerical model that takes into account the choice of the polarization of the probe mode and the corresponding coupling to the relevant Zeeman sublevels for each hyperfine state within the $5P_{3/2}$ manifold has to be employed.  We also assume that the intermediate state decays equally, via spontaneous emission, to both ground states; again, in practice, a more detailed numerical model of spontaneous emission from all Zeeman sublevels have to be taken into account.  In order to avoid the variation in the probe intensity that occurs in a standing wave cavity, it would be necessary to use a linear cavity consisting of three mirrors, as shown in figure 12 of Ref. ~\cite{Sarkar3}, with one of the mirrors being a perfect reflector, while the other two each being a partial reflector.  However, for simplicity of analysis, in what follows, we consider a two-mirror cavity, with a length of $L$ meter, an effective mode area of $A$, and a reflectivity of R for each mirror.  The input-output relation for such a cavity can be expressed as ~\cite{Gardiner, Aspelmeyer}:
\begin{equation}
\dot {\hat {\tilde a}} = -\frac {\kappa}2\ \hat {\tilde a} + i \delta \hat {\tilde a} + \sqrt {\kappa_{ex}} \xi + \sqrt {\kappa_o} \hat {\tilde f}
\label{eq:10}
\end{equation}
Here, the detuning is defined as the difference between the probe frequency ($\omega_p$) and the cavity resonance frequency ($\omega_c$): $\delta=\omega_p-\omega_c$.  The input probe field is assumed to be classical, defined as   $\alpha_{in}=\xi $ exp(-i$\omega_p$t),  and a mean value of $\abs {\xi}=\sqrt {N_{in}}$, where $N_{in}$ is the number of photons incident on the cavity in one second. The slowly varying amplitude of the field inside the cavity, transformed to a frame rotating at the frequency of the probe, is defined as $\hat {\tilde a}$=$\hat {a}$ exp(i$\omega_p$t).  The rate of decay of the intracavity intensity through the input mirror is defined as $\kappa_{ex}$, and any additional decay (including the decay through the output mirrror) is defined as $\kappa_o$, so that the net rate of decay is $\kappa=\kappa_{ex}+\kappa_o$.    Finally, the Langevin force operator in the last term of Eqn.~\ref{eq:10} obeys the relations $[\hat {\tilde f} (t),  \hat {\tilde f} ^\dagger (t')]=\delta (t-t')$ and $<\hat {\tilde f}>=0$.  In what follows, we assume that $\kappa_{ex}=\kappa_o=\kappa / 2$, since both mirrors have the same transimittivity, and other potential losses are ignored.  

The Hamiltonian for the whole system, including the atoms, the probe field inside the cavity, and the interaction among them, can be written as :
\begin{equation}
H=H_{cav}+H_{atm}+H_{int}+H_{src}
\label{eq:11}
\end{equation}
The four components of the Hamiltonian are defined as follows ~\cite{Aspelmeyer,Davis} (setting $\hbar=1$):
\begin{align}
& H_{cav} =\omega_c {\hat a}^\dagger {\hat a}  \nonumber \\
& H_{atm} = \sum_{j = 1}^{N} [2 \Delta {\op {\uparrow}}_j+(\omega_c+\Delta) {\op m}_j]  \nonumber \\
& H_{int} = \sum_{j = 1}^{N} [g \hat a {\op {m}{\uparrow}}_j+ g \hat a {\op {m}{\downarrow}}_j + h.c.]  \nonumber \\
& H_{src} = \sqrt {\kappa / 2} [ i\xi  {\hat a}^\dagger e ^ {-i \omega_p t}-i\xi ^ * \hat a e ^ {i \omega_p t}]
\label{eq:12}
\end{align}
where $g$ is the vacuum Rabi frequency for the cavit mode. 

The density operator, $\rho$, for the atoms and the cavity mode obey the following equation of motion:
\begin{equation}
\dot \rho =-  i  [H, \rho]+D(L_{cd})\rho + \sum_{j = 1}^{N}D(L_{\uparrow , j})\rho + \sum_{j = 1}^{N}D(L_{\downarrow , j})\rho
\label{eq:13}
\end{equation}
where we have defined: 
\begin{equation}
D(L)\rho = L \rho  L ^\dagger - \frac 1 2 \ \{L ^\dagger L, \rho \}
\label{eq:14}
\end{equation}
The Lindblad operator corresponding to cavity decay is $L_{cd}= \sqrt \kappa \hat a$ and those corresponding to spontaneous emission are $L_{\uparrow}=\sqrt { \Gamma / 2  } {\op {\uparrow}{m}}$ and $L_{\downarrow}=\sqrt {\Gamma / 2 } {\op {\downarrow}{m}}$.
 
Coherent excitation of the atoms only populates the $(N+1)$ symmetric collective states~\cite{Dicke, Arecchi, Sarkar2}. However, the total number of collective states, which include the asymmetric ones, is $2^{N}$, the size of the Hilbert space for $N$ two-level atoms~\cite{Sarkar2}. All of these states must be taken into account when considering the effect of spontaenous emission, which can couple to both symmetric and asymmetric states. Thus, even for a modest number of $N$ that would be relevant for a Schr\"odinger cat atomic interferometer, such an analysis is intractable.  As such, we will account for the effect of spontaneous emission heuristically, and exclude the Lindblad operators corresponding to spontaneous emission (this is the same approach used, for example, in Ref.~\cite{Davis}).  

Since the probe is highly detuned with respect to the two legs of the $\Lambda$ transition in Fig.~\ref{fig:11}, the intermediate state, $\ket m$, can be eliminated adiabatically, using the approach developed in Ref.~\cite{Reiter}.  The resulting Hamiltonian then can be expressed as :
\begin{equation}
H =( \omega_c + \epsilon {\hat J}_z ){\hat a}^\dagger \hat a +2 \Delta {\hat J}_z + \sqrt {\kappa / 2} [ i\xi  {\hat a}^\dagger e ^ {-i \omega_p t}-i\xi ^ * \hat a e ^ {i \omega_p t}]
\label{eq:15}
\end{equation}
where $\epsilon=2 g^2 / \Delta$ is the difference between the single-photon induced light-shifts experienced by the spin-up and spin-down states, which are equal and opposite in sign.  We now transform into a frame rotating at $H_A \equiv (\omega_p {\hat a}^\dagger \hat a + 2 \Delta {\hat J}_z$), which results in the following form of the Hamiltonian:
\begin{equation}
 H =(-\delta+ \epsilon {\hat J}_z ){\hat a}^\dagger \hat a +  \sqrt {\kappa / 2} [ i\xi  {\hat a}^\dagger -i\xi ^ * \hat a ]
\label{eq:16}
\end{equation}
It is now evident from the term in the first bracket of Eqn.~\ref{eq:16} that the detuning of the probe away from cavity resonance is modified by the light shift of the atoms.  The Lindblad operator for cavity decay under this transformation picks up a time-dependent phase factor: $L_{cd}= \sqrt \kappa \hat a$ exp $(-i \omega_p t)$. However, since the phase factor does not change $D(L_{cd})\rho$ (see Eqn.~\ref{eq:14}), we can write that effectively $L_{cd}= \sqrt \kappa \hat a$.  

Next, we assume that the intracavity field can be treated as the sum of a classical field and a weak quantum field:  $\hat a=(\alpha +\hat q)$, where $<\hat q>=0$.  We define the classical part as $\alpha=\zeta $ exp $(-i \omega_p t)$, so that $<\hat {\tilde a}>=\zeta$.  The steady state solution of Eqn. ~\ref{eq:10} then yields (with $\kappa_{ex}=\kappa / 2$):
\begin{equation}
\zeta = \sqrt { \frac \kappa 2 \ }  \frac \xi {(\kappa / 2 - i \delta})  \ 
\label{eq:17}
\end{equation}

The Hamiltonian now can be written as $H=H_a+H_b+H_c$, where:
\begin{align}
& H_a = -\delta {\hat q}^\dagger \hat q + \epsilon J_z [{\abs {\alpha}}^2 + {\hat q}^\dagger \hat q + \alpha^*  \hat q + \alpha {\hat q}^ \dagger]\nonumber \\
& H_b = -\delta {\abs {\zeta}}^2+ {\abs {\xi}}^2 \sqrt {\kappa /2} [\frac {2 \delta} {\delta^2 + \kappa^2 /4}] \ \nonumber \\
& H_c = i(\kappa / 2) [\alpha {\hat q}^ \dagger-\alpha^*  \hat q ]
\label{eq:18}
\end{align}
and the cavity-decay Lindblad operator becomes $L_{cd}=\sqrt \kappa (\alpha + \hat q)$.  Since $H_b$ does not involve any operators, it can be transformed out trivially, yielding $H=H_a+H_c$ and  $L_{cd}=\sqrt \kappa (\alpha + \hat q)$.  Furthermore, it can be shown easily that the equation of motion for the density matrix remains unchanged when this combination of the Hamiltonian and the cavity-decay Lindbald operator is replaced by the combination of $H=H_a$ and  $L_{cd}=\sqrt \kappa \hat q$.  

Adiabatic elimination of the weak cavity mode, $\hat q$, again using the approach developed in Ref.~\cite{Reiter}, yields the following expressions for the effective Hamiltonian and the Linblad operator for the spin dynamics:
\begin{align}
& H = \chi J^2_z \nonumber \\
& L = \sqrt {(2 \chi / \tilde \delta)} J_z
\label{eq:19}
\end{align}
with the squeezing parameter, $\chi$, given by:
\begin{equation}
\chi = \tilde \delta {(1+{\tilde \delta}^2 )}^{-2} {\abs {\xi}}^2 {\tilde \epsilon}^2
\label{eq:20}
\end{equation}
where we have defined the probe detuning (away from cavity resonance) normalized to the half-width of the cavity resonance as $\tilde \delta \equiv \delta / (\kappa /2)$, and the single photon induced differential light shift for each atom normalized to the half-width of the cavity resonance as $\tilde \epsilon \equiv \epsilon / (\kappa /2)$.  Finally, as noted earlier, the quantity ${\abs {\xi}}^2$ represents the number of photons incident on the cavity per second.  

An important factor that determines the degree of fidelity achievable in the squeezing process is the single-atom cooperativity, defined as $\mathscr {C} \equiv 4 g^2 / (\kappa \Gamma)$.  In terms of this factor, the squeezing parameter can be expressed as:
\begin{equation}
\chi = \tilde \delta {(1+{\tilde \delta}^2 )}^{-2} {\abs {\xi}}^2 {\mathscr {C}}^2 {(\Gamma / \Delta)}^2
\label{eq:21}
\end{equation}

As an example, consider a linear cavity with length $L$, effective mode area of $A$, and a transmittivity of $T$ for each end mirror.  We then find that $\mathscr {C}=\mathscr {A}/ (AT)$, where $\mathscr {A}=8 \pi \hbar \omega_p \Gamma / I _{SAT}$, with $I_{SAT}$ being the saturation intensity for each leg of the $\Lambda$ transition.  For $^{87}Rb$ atoms, assuming that $I_{SAT}$ is twice that of the cycling transition, we get $\mathscr {A} \approx 3.6*10^{-12} m^2$.  For a mirror with a reflectivity of 99.999\%, so that $T=T_o=10^{-5}$, and mode area of $(20*10^{-6}m)^2$, we get $\mathscr {C} \approx 900$.  If we define the mode area to be $D^2$, and a reference value of $D$ to be $D_o = 20*10^{-6}m$, then we get $\mathscr {C} \approx 900*{(D_o/D)}^2 (T_o/T)$.  If we denote as $P$ the incident power, a reference power of $P_o=10^{-3}$ Watt, and a reference normalized detuning of ${\tilde \delta}_o=10^2$, we then get, in units of ${sec}^{-1}$:
\begin{equation}
\chi \approx 10^8 {({{\tilde \delta}_o} / {\tilde \delta} )}^3 {(P/{P_o})}^2 {(D_o/D)}^4{(T_o/T)}^2
\label{eq:22}
\end{equation}
Thus, for $\tilde \delta = {\tilde \delta}_0$, $P=P_o$, $D=D_o$ and $T=T_o$, the time needed for producing the SC state would be $t_{SC} \equiv \pi / 2 \chi \approx 15$ nsec.  For more moderate choice of parameters, e.g. $D=10D_o$ and $T=10T_o$, we have $\mathscr {C} \approx 0.9$, and $t_{SC}  \approx 15   \mu$sec.  If we increase the power to $P=10P_o=10^{-2} $ Watt, which is still very modest, we get $t_{SC}  \approx 0.15   \mu$sec.  

We now consider the effect of the dephasing due to the cavity decay, as well as spontaneous emission.   In order to express the results quantitatively, we note first that while the actual improvement in the performance of the interferometer or the clock is given by $(\Lambda / \Lambda_{SQL})$, it is customary in the literature to quote the value of  ${(\Lambda / \Lambda_{SQL})}^2$.  To remain consistent with this custom, we define $\mathscr {F}$ as the factor of improvement, as follows:
\begin{equation}
\mathscr {F} \equiv {\qty( \frac {\Lambda} {\Lambda_{SQL}} \ )}^2 = \frac {{({\Delta \phi}_{SQL})}^2} {{( \Delta \phi )}^2}  \ = \frac {(1/N)} {{( \Delta \phi )}^2}  \
\label{eq:23}
\end{equation}

To incorporate the effects of cavity decay and spontaneous emission, we can write:
\begin{equation}
{( \Delta \phi )}^2 =  {({\Delta \phi}_{COH})} ^2 +  {({\Delta \phi}_{CAV})} ^2 + {({\Delta \phi}_{SE})} ^2   
\label{eq:24}
\end{equation}
where ${\Delta \phi}_{COH}$ represents the phase variance due to the coherent evolution of the spins,  ${\Delta \phi}_{CAV}$ represents the phase variance due to cavity decay, and  ${\Delta \phi}_{SE}$ represents the phase variance due to spontaneous emission.  Thus, we get
\begin{equation}
\mathscr {F} = {\qty [ N  \qty{ {({\Delta \phi}_{COH})} ^2 +  {({\Delta \phi}_{CAV})} ^2 + {({\Delta \phi}_{SE})} ^2 }  ] }^{-1} 
\label{eq:25}
\end{equation}

We also deifne ${\Delta \phi}_{IDL}$ to be the phase variance under ideal conditions (i.e., when there is no cavity decay or spontaneous emission).  We recall that, under this condition, the signal for the CD-SCAIN and CD-SCAC is $S_{IDL}=\ev*{{\hat J}_z }=(-N/2) Cos(N \phi)$, with ${\Delta S}_{IDL}=(N/2) \abs {Sin(N \phi)}$ and ${(\partial_\phi S)}_{IDL}=(N^2/2) Sin(N \phi)$.  Thus, in this case
\begin{equation}
{\Delta \phi}_{COH} = {\Delta \phi}_{IDL}= \frac {{\Delta S}_{IDL}} {{\abs {\partial_{\phi} S}}_{IDL} } \ = \frac {1} {N} \
\label{eq:26}
\end{equation}
independent of the value of $\phi $, so that  ${\mathscr {F}}_{IDL}=N$. Of course, in general, ${\Delta \phi}_{COH} \neq {\Delta \phi}_{IDL}$.

For the general case where some of the spins are dephased due to either cavity decay or spontaneous emission, the number of atoms, defined as $\tilde N$, that will constitute the SC state, representing coherent evolution of the spins, will be less than $N$.  Thus, we can write the coherent part of the signal, its standard deviation and its phase gradients as:

\begin{align}
&\tilde S=(-\tilde N/2)Cos(\tilde N \phi) \nonumber \\
&\Delta \tilde S=(\tilde N/2) \abs { Sin(\tilde N \phi) } \nonumber \\
&\abs {\partial_{\phi} {\tilde S}}= ({\tilde N}^2 /2) \abs { Sin(\tilde N \phi) }
\label{eq:27}
\end{align}
The angular variances are determined by the ratios of the signal variances and the phase gradient of the coherent part of the signal, as follows:
\begin{align}
&{(\Delta \phi_{COH})}^2={(\Delta \tilde S)}^2 /{(\partial_{\phi} {\tilde S})}^2  \nonumber \\
&{(\Delta \phi_{CAV})}^2={({\Delta  S}_{CAV})}^2 /{(\partial_{\phi} {\tilde S})}^2 \nonumber \\
&{(\Delta \phi_{SE})}^2={({\Delta  S}_{SE})}^2 /{(\partial_{\phi} {\tilde S})}^2
\label{eq:28}
\end{align}
As illustrated in Fig. 2 of Ref.~\cite{Itano}, during the operation of a clock or an interferometer, one measures the signal at two different phases, ($\phi - \delta \phi$) and ($\phi + \delta \phi$), and $\phi$ is varied until these two signals are equal.  The value of $\phi$ determined this way correspond to the value for which the signal is maximum.  Deviation of this value of $\phi$ from the quiescent value (typically zero) is then used to determined the amount of rotation in the case of an atom interferometric gyroscope (or the frequency deviation in the case of a clock).  For the CD-SCAIN or the CD-SCAC, a convenient value of $\delta \phi$ is $\pi/(2 \tilde N)$.  This corresponds to making measurements at the point of the signal fringe where both the magnitude of the phase gradient of the coherent signal ($\abs {\partial_{\phi} \tilde S} $) and the standard deviation of the coherent signal ($\Delta \tilde S$) have their maximum values, as can be seen from Eqn.~\ref{eq:27}.  Thus, in what follows, all quantities in  Eqn.~\ref{eq:28} are to be evaluated at $\phi_o=\pi/(2 \tilde N)$.  We thus get, 
\begin{equation}
\mathscr {F} = {\qty [ \frac {4N} {{\tilde N}^4} \  \qty{ \frac {{\tilde N}^2} {4} \ +  {({\Delta S}_{CAV})} ^2 + {({\Delta S}_{SE})} ^2 }  ] }^{-1} 
\label{eq:29}
\end{equation}
It should be noted that the effect of the variances due to cavity decay as well due to spontaneous emission are strongly suppressed because of the magnified phase gradient of the signal.  This is yet another manifestation of the robustness of the CD-SCAIN and the CD-SCAC.  
	
We denote $\delta N \equiv N - \tilde N$ as the reduction in the peak-to-peak amplitude of the signal.  As we show next, the cavity decay process and the spontaneous emission process both contribute to $\delta N$, in addition to producing additional variances in the signal.  The net reduction in the value of $\mathscr {F}$ is due to a combination of these factors.  In what follows, we now estimate the values of ${\Delta S}_{CAV}$, ${\Delta S}_{SE}$, and $\delta N$ resulting from these processes, in order to determine the value of $\mathscr {F}$.

Consider first the effect of cavity decay.  A rigorous study of the effect of cavity decay on the SCAIN protocol would require carrying out the whole analysis using the density matrix approach, based on the Hamiltonian and the Lindblad operator shown earlier in  Eqn.~\ref{eq:19}.  For N atoms, the size of the density matrix will be $N^2$.  To solve the equation of motion, one has to form a vector consisting of all the elements of the density matrix, and the propagator matrix that determines the time derivative of this vector would have dimensions of $N^2XN^2$ ~\cite{Shahriar4}.  This is a daunting task for a large value of N.  For example, to determine the time evolution for $N=10^3$, one has to diagonalize a matrix with $10^{12}$ elements.  Diagonalization is necessary even if one wants to make use of direct numerical integration, since the smallest time steps to be used for the integration must be significantly smaller than the inverse of the largest eigenvalue of the propagator matrix, and the duration of the evolution must be significantly larger than the inverse of the smallest eigenvalue.   In the near future, we will carry out such an analysis, for as large a value of N as feasible, within the constraint of computations resources.  For the present work, we make use of a perturbative approach consisting of two steps.  In the first step, we ignore the effect of the cavity decay, and use the Hamiltonian of Eqn.~\ref{eq:19} to evolve the quantum state of the ensemble coherently.  The results of this step have already been documented in the preceding sections.  In the second step, we estimate the effect of the cavity decay by consider the density matrix equation of motion attributable to only the Lindblad operator in Eqn.~\ref{eq:19}.  Such a two-step approach has also been used in Ref. ~\cite{Davis}, for example.  In what follows we carry out the second step of this analysis.

We define $\gamma \equiv 2 \chi / {\tilde \delta}$, so that the Lindblad operator in  Eqn.~\ref{eq:19} can be expressed as $L=\sqrt \gamma J_z$.  From Eqn.~\ref{eq:14}, it then follows that the incoherent part of the evolution of the density matrix can be written as:
\begin{equation}
\dot \rho = \gamma J_z \rho J_z^\dagger  - \frac \gamma 2 \ \{J_z^\dagger J_z, \rho \}
\label{eq:30}
\end{equation}
Using the fact that $\ev* { \dot {\hat O} } =tr(\dot \rho \hat O)$ for any operator $\hat O$,  it can be shown that:
\begin{align}
& \ev* { {\dot J}_\pm } =  -(\gamma /2) \ev* {J_\pm}   \nonumber \\
& \ev* { {\dot J}_x^2 } = - \gamma \ev* { J_x^2 } + \gamma \ev* { J_y^2 };  \ev* { {\dot J}_y^2 } = - \gamma \ev* { J_y^2 } + \gamma \ev* { J_x^2 } \nonumber \\
& \ev* { {\dot J}_z } = 0; \ev* { {\dot J}_z^2 } = 0
\label{eq:31}
\end{align}
where $J_\pm \equiv J_x+iJ_y$, so that  $\ev* { {\dot J}_x } =  -(\gamma /2) \ev* {J_x}$ and $\ev* { {\dot J}_y } =  -(\gamma /2) \ev* {J_y}$.   In the protocol for creating the CD-SCAIN and the CD-SCAC, we have the following values at the beginning of the OATS process: $\ev*{J_x}=\ev*{J_z}=0$, $\ev*{J_y}=J$, $\ev*{J_x^2}=\ev*{J_z^2}=J/2$, and $\ev*{J_y^2}=J^2$.    For $\gamma t \ll 1$, we then find that, at the end of the OATS process, we have (keeping in mind that this evolution is due to the cavity decay effect only): $\ev*{J_x}=\ev*{J_z}=0$, $\ev*{J_y}\approx J (1-\gamma t/2)$, $\ev*{J_z^2}=J/2$, and $\ev*{J_x^2}=(J/2)(1+2J \gamma t)$ and $\ev*{J_y^2}=J^2(1-\gamma t)$, leaving the value of $\ev*{{\bold J}^2}=J(J+1)$ unchanged.  Thus, we get:
\begin{align}
& {(\Delta J_x)}^2=\ev* {J_x^2}-{\ev* {J_x}}^2 =J/2+J^2 \gamma t  \nonumber \\
& {(\Delta J_y)}^2=\ev* {J_y^2}-{\ev* {J_y}}^2 =0 \nonumber \\
& {(\Delta J_z)}^2=\ev* {J_z^2}-{\ev* {J_z}}^2 =0
\label{eq:32}
\end{align}
Therefore, the net effect of the cavity decay is the increase in the value of  ${(\Delta J_x)}^2$ by an amount $J^2 \gamma t$, and the decrease in the length of $\ev* {J_y}$ by $J \gamma t/2$.  We recall that, in the protocol for the CD-SCAIN and the CD-SCAC, the auxiliary rotation immediately after the OATS process maps the y-component of the spins to the z-component.  Thus, the reduction in the length of $\ev* {J_y}$ maps to a reduction in the length of $\ev* {J_z}$, and therefore a reduction in the coherent signal.   On the other hand, the increase in the variance of $J_x$ does not contribute directly to an increase in the variance of the signal (which corresponds to measuring $J_z$).  However, we allow, as an upper limit, this increase in variance of $J_x$ as a corresponding increase in the variance of the signal.  Next, note that, under ideal conditions, the initital conditions for the inverse OATS process, for both CD-SCAIN and the CD-SCAC, at $\phi= \pm \pi/(2N)$ (which is the value of the phase at which measurements are to be made, as discussed earlier), the initial conditions are the same as those at the beginning of the OATS process, as can be seen from Figs.~\ref{fig:2} and ~\ref{fig:8}.  Thus, we can assume that the effects of cavity decay for the inverse OATS process are essentially the same as what we estimated above for the OATS process.  As such, we get, for the OATS and the inverse OATS process combined: 

\begin{align}
& {(\Delta S_{CAV})}^2=2*J^2 \gamma t =N^2  \gamma t / 2 \nonumber \\
& \delta N_{CAV}=2* J \gamma t/2= N\gamma t/2
\label{eq:33}
\end{align}
Next, we consider the effect of spontaneous emission.  As noted earlier (see the paragrpah after Eqn.~\ref{eq:14}), for a large value of N, it is virtually impossible to account for the effect of spontaneous emission in an analytic manner.  As such, we account for this in a heuristic manner, similar to what is done in Ref. ~\cite {Monika}.  The number of photons in the cavity is $\zeta ^2$, where $\zeta$ is given by Eqn.~\ref {eq:17}.   Assuming that the atomic excited state ($\ket {m}$ in Fig.~\ref{fig:11}) decays at the rate of $\Gamma$, and that $\Delta \gg \Gamma$ as well as $\Delta \gg g$, the number of photons scattered by each atom happens at the rate of:
\begin{equation}
\tilde \Gamma={(g/\Delta)}^2 {\abs {\zeta}}^2 \Gamma=\frac {\chi}{2 \mathscr {C}} \ \frac {(1+{\tilde \delta}^2)} {\abs {\tilde \delta}} \
\label{eq:34}
\end{equation}
Spontaneous emission causes the spin to flip randomly, from up to down and vice versa.    Thus, the value of  $J_z$ decreases via random walk as:
\begin{equation}
{(\Delta J^{SE}_z)}^2 \approx \mathscr {P} (1/2) N  \tilde \Gamma t
\label{eq:35}
\end{equation}
Here, $\mathscr {P}$ is the probability of spin flip, which is 1/2 for a symmetric system, such as the one depicted in Fig.~\ref{fig:11}, so that we get, in the limit of $\abs{\tilde \delta} \ll 1$ :
\begin{equation}
{(\Delta J^{SE}_z)}^2 \approx  \frac {N \chi t}{8 \mathscr {C}} \ \abs {\tilde \delta} 
\label{eq:36}
\end{equation}
Thus, we can now write that:
\begin{align}
& {(\Delta S_{SE})}^2= \frac {N \chi t}{8 \mathscr {C}} \ \abs {\tilde \delta}  \nonumber \\
& \delta N_{SE}=\sqrt { \frac {N \chi t}{8 \mathscr {C}} \ \abs {\tilde \delta} }
\label{eq:37}
\end{align}
We define $\Theta \equiv \delta N /N$, where $\delta N=\delta N_{CAV}+\delta N_{SE}$, so that  $\tilde N=N (1-\Theta)$.  We also specify that $\chi t=\pi/2$ as the condition for creating the SC state, and assume that $\Theta \ll 1$.  Inserting Eqns. ~\ref{eq:33} and ~\ref{eq:37} in Eqn. ~\ref{eq:29}, we then get:
\begin{equation}
\mathscr {F} \approx {\qty [ \frac {1}{N} \ (1+2\Theta) +  \frac {2 \pi}{N^2} \ (1+4\Theta) \qty{ \frac {N}{\abs {\tilde \delta}} \ + \frac {\abs {\tilde \delta}}{8 \mathscr {C}} \ }  ] }^{-1} 
\label{eq:38}
\end{equation}
%
\begin{figure}[h]
\includegraphics[scale=0.35]{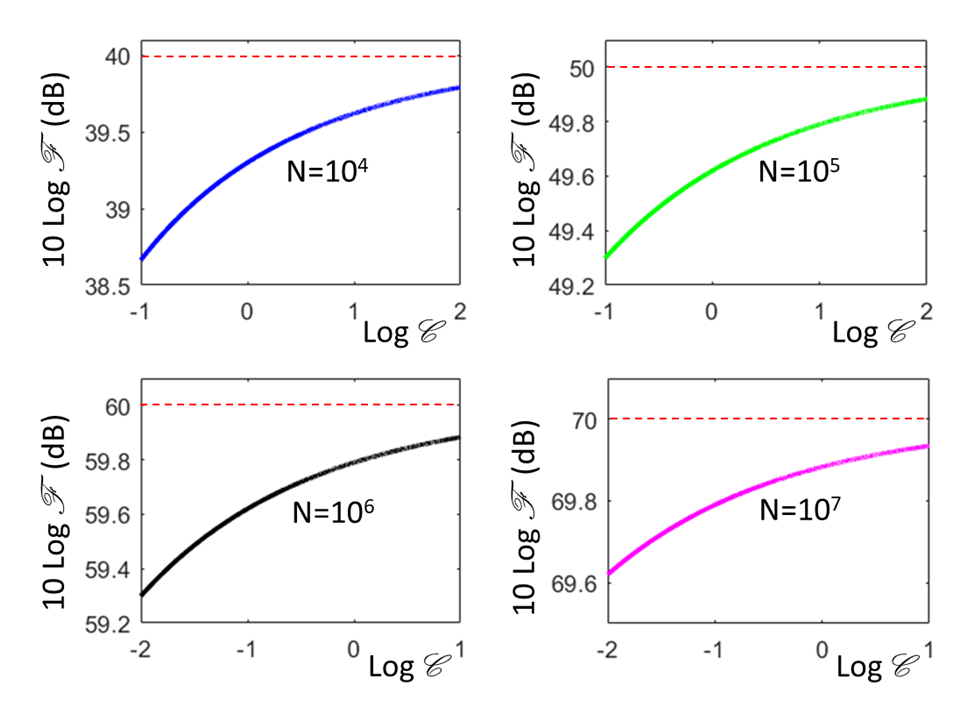}
\caption{Illustration of the factor of improvement, $\mathscr {F}$, as a function of the cooperativity parameter, $\mathscr {C}$, for four different values of $N$.  In each case, the ideal value of  $\mathscr {F}$ is indicated by the dotted red line.}
\label{fig:12}
\end{figure}

The term inside the curly brackets in Eqn.~\ref{eq:38} is minimized for $\abs {\tilde \delta}=\sqrt {8{\mathscr {C}}_N}$, where ${\mathscr {C}}_N \equiv N \mathscr {C}$ is the collective cooperativity parameter.  For ${\mathscr {C}}_N \gg 1$, and this value of $\abs {\tilde \delta}$, we get:

\begin{align}
& \Theta \approx {[\pi^2/(32{\mathscr {C}}_N )]}^{1/4}  \nonumber \\
& \mathscr {F} \approx {\qty [ \frac {1}{N} \ (1+2\Theta + 8\Theta^2+32\Theta^3 )   ] }^{-1} 
\label{eq:39}
\end{align}
Since $\Theta \ll 1$ for ${\mathscr {C}}_N \gg 1$, we thus get  $\mathscr {F} \approx N (1-2\Theta) $.

In Figure ~\ref{fig:12}, we have illustrated the factor of improvement, $\mathscr {F}$, as a function of the cooperativity parameter, $\mathscr {C}$, for four different values of $N$.  In each case, the ideal value of  $\mathscr {F}$ is indicated by the dotted red line.   As can be seen, even for $\mathscr {C}=0.01$, which should be easily accessible, based on the analysis shown after Eqn. ~\ref{eq:21}, the achievable value of $\mathscr {F}$ is within less than 1 dB of the maximum possible value of 70 dB for ten million atoms.  

In the preceding discussion, we have addressed the effect of residual spontaneous emission heuristically.   This model may not account fully for the deleterious effects of spontaneous emission.  Consider, for example, a situation where the centers of mass of the two components of the SC state are separated spatially by a distance  $\mathscr {D}$ in the z-direction.  If a spontaneous emission event occurs, we will call the event "distinguishable" if, in principle, it is possible to determine which component of the SC state produced the photon; otherwise, we will call the event "indistinguishable." If the emission is distinguishable, then the SC state would collapse to a single collective state, and there would be no interference ~\cite {Yuping, Pawlowski}.  The distinguishablity of the emission event will be determined by the size of $\mathscr {D}$.  The relevant length scale here is the wavelength of the emitted photon: $\lambda_P$.   For the D2 transition in Rb, $\lambda_P \approx 780$ nm.  Indistinguishability would only hold for $\mathscr {D} < \lambda_P$ ~\cite {Dicke, Romero}.  We also note here that the dephasing caused by cavity decay is also "indistinguishable," since the cavity mode is much larger than the separation between the two components of the SC state.  

For the CD-SCAC, as well as variations of the CD-SCAC protocol for magnetometry or nuclear magnetic resonance, this condition can be easily satisfied, since the recoil corresponding to a microwave transition is very small.  For example, for the hyperfine transition in the ground state of $^{87}Rb$, the recoil velocity is $\approx 0.2 \mu$m/sec.  As shown earlier, a typical time duration for the squeezing process to produce the SC state is $\approx 0.15 \mu$sec.  Thus, during the squeezing process, the value of $\mathscr {D}$ would be only $\approx 0.03$ pico meter, far less than $\lambda_P$.   Consider next the case of the CD-SCAIN, for which the recoil velocity would be $\approx 12$ mm/sec.  At the point of maximum separation between the two arms, the condition of  $\mathscr {D} \approx \lambda_P$ would be reached for a dark zone duration of $\approx 65 \mu$ sec.  For a typical dark zone duration used for accelerometry or rotation sensing, which is much longer than this, the value of $\mathscr {D}$ would far exceed $\lambda_P$.  However, it should be noted that the squeezing process is carried out at the onset of the splitting, and the unsqueezing happens after the two paths have come back to each other (see Fig ~\ref{fig:2}).  Thus, during the squeezing/unsqueezing steps, each with a typical duration of $\approx 0.15 \mu$sec, the value of $\mathscr {D}$ would be only $\approx 1.8 $nm, which is much smaller than $\lambda_P$.  As such, for both the CD-SCAC and the CD-SCAIN, the spontaneous emission events would not lead to collapse of the SC state to a single collective state.  

One must also take into account the possibility of spontaneous emission during the pulses other than those used for squeezing and unsqueezing.  For the CD-SCAC, this is not an issue, since these pulses would employ microwaves.  For the CD-SCAIN, this is important for the $\pi$-pulse that reverses the direction of motion for the two arms, since the separation between the two components of the SC state would typically be much larger than the wavelength of light at this point, as noted above.  However, by using strong laser fields and large detunings for each leg of the Raman transition, it should be possible to suppress the probability of a spontaneous emission during this pulse to be adequately small.   

We recall that, for the CD-SCAIN and the CD-SCAC, cases with only one parity (even N) contributes the signal with N-fold magnified fringes, while cases with the other parity (odd N) contributes essentially zero signal.  While cavity decay (which is "indistinguishable," as noted above) and the "indistinguishable" spontaneous emissin events would not cause a collapse of the SC state, these would change the parity of the coherent ensemble, defined as the atoms that contribute to the coherent part of the signal (as defined in eqn. ~\ref{eq:27}).   The final parity of the coherent ensemble at the end of the protocol would determine whether it would contribute to the N-fold magnified fringe signal, or the essentially zero signal.  

Finally, we consider the issue of potential loss of particles due to collisions.  As discussed, for example, in Refs. ~\cite {Yuping} and ~\cite {Pawlowski}, this can be of potentially significant concern in creating SC states of Bose-condensed atoms.  However, for the systems being considered here, the density of atoms is small enough to ignore collisions among the atoms in the SC state.  Collisions with background atoms can also be made negligible by using ultra-high vacuums produced under cryogenic conditions ~\cite {Willems, Alpha}.  For example, in Refs. ~\cite {Romero, Romero2, Romero3}, which address this issue in the context of attempts to create macroscopic superposition of nanoparticles, it has been shown that collisions with background atoms become negligible for a vacuum of $\approx 10^{-16}$ Torr.  Such pressures have been previously realized in cryogenic environments ~\cite{Gabrielse}.   

\section{Discussion of Closely Related Work}
\label{Section 6}
The basic concept underlying the protocol described in this paper falls within the broad category of so-called interaction based readout (IBR) schemes ~\cite{Pezze}.  A systematic discussion of the IBR schemes can be found, for example, in references ~\cite{Haine1}.   In this reference, many different versions of IBD schemes are discussed, including one that is similar to the protocol described here.  In another paper ~\cite{Huang}, the authors analyze an IBR scheme that is also similar to our protocol.  In reference ~\cite{Haine3}, a systematic study is carried out to determine the maximum possible robustness against excess noise.   This study concludes that our protocol, along with similar ones presented in references ~\cite{Haine1} and ~\cite{Huang}, achieve the maximum possible sensitivity against excess noise.  The optimal robustness against excess noise for a different IBR scheme, employing twist-and-turn entanglement, is investigated in reference ~\cite{Haine2}.   Indeed,  references ~\cite{Haine2}, ~\cite{Haine3} and ~\cite{Huang} have cited the arxiv version of our protocol ~\cite{Fang}.  It has been noted in reference ~\cite{Haine3} that the robustness of our protocol against excess noise is ostensibly stronger than that of  the corresponding protocol in reference ~\cite{Haine1}, attributing the difference to the fact that the noise model used by us (which is akin to the one used in reference ~\cite{Davis}) is different from the ones used in references ~\cite{Haine1}.  Similarly, the roubustness of our protocol is ostensible stronger than that of the protocol in reference ~\cite{Huang}, for the same reason.   While the maximum robustness protocols presented in references ~\cite{Haine1} and ~\cite{Huang} are similar to ours, we note that the details of our protocol contain significant differences and augmentations, as discussed next.

An important aspect of what we describe in this paper is the challenge in implementing this protocol for large number of cold atoms released from a magneto-optic trap.  For such a scenario, the parity of the number of atoms, $N$ (i.e., whether $N$ is even or odd) is not known.  On the other hand, the protocol produces very different signals for even versus odd values of $N$.  What we show is that one can choose to operate  the protocol designed, for example, for even values of $N$.  For instances where $N$ is odd, this protocol will produce a vanishing signal, so that, when averaged over many instances, the net signal would simply correspond to running the protocol for even values of $N$ only, with the effective number of $N$ being reduced by a factor of 2.  Reference ~\cite{Haine3} notes that different protocols are necessary for odd versus even values of N.  However, it does not show what happens if a fixed protocol is used, while the system has randomly occurring odd and even values of N.  Reference ~\cite{Haine1} and ~\cite{Huang} do not consider this issue at all. 

Another important issue addressed in this paper is the application of the protocol to an atomic interferometer where the trajectories of the two paths get physically separated spatially, thus requiring the application of an additional $\pi$ pulse.  We have considered the application of the protocol to this case explicitly, and have contrasted its behavior with the case of an atomic clock, where this extra $\pi$ pulse is not required.   Even for the case of the atomic clock, there are some difference in the details of the steps in our protocol as compared to those in references ~\cite{Haine1} and ~\cite{Huang}.  Specifically, in our protocol, we apply a rotation immediately after the initial one axis twist squeezing that produces the cat state, but before the phase accumulation starts.  This rotation is undone before the application of the unsqueezing pulse prior to detection.  The scheme in reference ~\cite{Haine1} does not employ this additional pair of pulses.  In reference ~\cite{Huang}, two such pulses are applied, but both in the same direction: one immediately before the unsqueezing pulse, and one immediately after the unsqueezing pulse; this is different from our sequence of pulses.  We believe these differences are attributable in part to different assumptions about initial states and the axis of rotation due to accumulation of phase difference.  It should also be noted that in references ~\cite{Haine1} and ~\cite{Haine3}, the detection requires measurement of the population of all collective states.  This is distinctly different from our case, where, for the case that produces the maximal robustness against excess noise, we measure the mean value of all the pseudospins in the z-direction (a process denoted as  conventional detection in this paper).

In references ~\cite{Haine1}, ~\cite{Haine3} and ~\cite{Huang}, the robustness of the protocol against excess noise is explained by showing how the Fisher information is influenced in the presence of excess noise.  While this is certainly correct, it does not seem to provide a simple and transparent reason for the robustness.  In contrast, we have in this paper offered a simple explanation of the robustness, which is as follows.  In this protocol, the signal fringes are amplified by a factor of $N$, while the quantum projection noise is magnified by a factor of $\sqrt{N}$ more than the noise under the standard quantum limit, reaching a value of $N$.  Thus, it is clear that the classical noise would only become relevant (reducing the sensitivity by a factor of $\sqrt{2}$) when it is very large, namely equaling $N$.  

We also show explicitly how the measurement basis affects very strongly the robustness against excess noise.  Specifically, we show that if one chooses to measure the population of one of the extremal collective states, the process is as sensitive to classical noise as conventional techniques (such a two axes counter twist squeezing), so that a classical noise of unity would reduce the sensitivity by a factor of $\sqrt{2}$.  In contrast, when the mean value of all the pseudospins in the z-direction is measured, the excess noise has to be very large (namely $N$) for the sensitivity to drop by a factor of $\sqrt{2}$.

Most importantly, in this paper we have addressed, in explicit and quantitative details, the critical question of the effect of dissipation during the OAT process employing an optical cavity.  Specifically, we have shown, employing the cavity input-output relation and the density matrix formulation involving Langevin noise operators, how the effect of the dissipative processes are strongly suppressed, due to the fact that the protocol entails phase magnification by a factor of $N$, and enhanced quantum noise by a factor of $\sqrt{N}$.  We thus find that the maximal achievable sensitivity is very close to the ideal limit, as shown in Fig. ~\ref{fig:12}.  This is to be contrasted with the findings of a similar analysis carried our in reference ~\cite{Davis}, for the echo squeezing protocol.  In that case, the dissipation during the cavity mediated OAT process limits the maximum achievable sensitivity to be far below the ideal value.  In reference ~\cite{Huang}, a brief discussions is presented regarding the effect of dissipation during the read-out process only.  However, this discussion is presented in the context of a generalized dissipation parameter, without describing the experimental conditions that would correspond to a given value of the dissipation parameter.  In contrast, we have considered explicit experimental parameters, such as the cavity cooperativity parameter, the number of atoms, and the laser power in the probe, and shown that sensitivity very close to the ideal limit is achievable for experimentally accessible values of these parameters.  References ~\cite{Haine1} and ~\cite{Haine3} do not address this issue of dissipation during the squeezing and unsqueezing processes.

\section{Conclusion}
\label{Section 7}
Atomic precision metrology is of importance for practical applications, such as time keeping, rotation sensing, accelerometry and magnetometry.  It also plays a key role in investigations of fundamental physics, including search for electron electric dipole moment, tests of general relativity, and detection of dark matter.  Under ideal conditions, the sensitivity of an atomic sensor is at the standard quantum limit, dictated by the quantum projection noise.  This limit can be circumvented by making use of entangled states of atoms.  In particular, the use of highly entangled states can enable one to reach the Heisenberg limit, which represents an improvement by a factor of $\sqrt{N}$, where $N$ is the number of atoms interrogated.  However, such a process is typically more sensitive to excess noise than conventional sensors.  Here, we describe a protocol for an atom interferometer that can reach the Heisenberg limit of sensitivity, while being more insensitive to excess noise than a conventional sensor.   Using spin-squeezing, the sensitivity can be increased, either by lowering the quantum noise or via phase amplification, or a combination thereof. In this paper, we have shown how to increase the sensitivity to the Heisenberg limit, while increasing the quantum noise by $\sqrt{N}$, thereby suppressing by the same factor the effect of excess noise. The protocol makes use of a Schr\"odinger Cat state representing a mesoscopic superposition of two collective states of $N$ atoms, behaving as a single entity with an $N$-fold increase in Compton frequency.  We show that the $N$-fold phase magnification can be produced under two different methods of detection: one where the population of an extremal collective state is measured, and the other where the mean value of the pseudo-spins of all atoms is measured.  However, the suppression of sensitivity to excess noise is achieved if the latter method is used, while the former method makes it extremely sensitive to excess noise.  We also show how to realize an atomic clock based on such a Schr\"odinger Cat state, with an $N$-fold increase in the effective transition frequency. We show how the signals, for a given protocol, produces drastically different signals for different parities of $N$, for both detection methods.  For a system, such as atoms released from a magneto-optic trap, that produces both odd and even values of $N$ with equal probability, we show that averaging over many instances filters out the signal for one parity, thus allowing a sensitivity that is within a factor of $\sqrt{2}$ of the Heisenberg limit, while maintaining the robustness against excess noise. We have shown both numerical and analytical results for the ideal behavior of the Schr\"odinger Cat state based atomic interferometer and atomic clock. We have also discussed potential experimental constraints for implementing this scheme, using one axis twist squeezing employing the cavity feedback scheme, and shown that the effects of cavity decay and spontaneous emission are highly suppressed due to the $N$-fold phase magnification. We have found that even for a modest value of the cavity cooperativity parameter, of the order of 0.01, which should be readily accessible experimentally, the maximum improvement in sensitivity can be very close to the ideal limit, for as many as ten million atoms.  We also discuss related protocols that have been proposed recently, and point out the similarities and differences of these with what is proposed here.  We believe that the concepts proposed here pave the way for realizing atomic interferometers and atomic clocks, as well as other atomic sensors based on excitation of pseudo-spins in an effective two level system, such as a magnetometer,  with sensitivity very close to the Heisenberg limit, without requiring rigorous suppression of excess noise, for a range of parameters that are rather easily accessible experimentally.   

Funding. National Science Fioundation (NSF)(DGE-$0801685$; DMR-$1121262$); Air Force Office of Scientific Research (AFOSR)(FA$9550$-$09$-$01$-$0652$)

Disclosure. The authors declare no conflict of interest.

\bibliographystyle{apsrev}

\end{document}